\documentclass[namedreferences]{solarphysics}
\usepackage[hyperref,optionalrh]{spr-sola-addons} 
\usepackage{graphicx}        
\usepackage{color}           
\usepackage{siunitx}

\newcommand{\aap}{    {\it Astron. Astrophys.}}

\newcommand{\apj}{    {\it Astrophys. J.}}
\newcommand{\apjl}{   {\it Astrophys. J. Lett.}}
\newcommand{\apjs}{   {\it Astrophys. J. Suppl.}}

\newcommand{\solphys}{{\it Solar Phys.}}
 
\newcommand{\ssr}{    {\it Space Sci. Rev.}} 

\begin{document}

\begin{article}
\begin{opening}

\title{Super-Resolution of SOHO/MDI Magnetograms of Solar Active Regions Using SDO/HMI Data 
and an Attention-Aided Convolutional Neural Network}

\author[addressref={aff1,aff2}]{\inits{C.}\fnm{Chunhui}~\lnm{Xu}}
\author[addressref={aff1,aff2},corref,email={wangj@njit.edu }]{\inits{J. T. L.}\fnm{Jason T. L.}~\lnm{Wang}}
\author[addressref={aff1,aff3,aff4},corref,email={haimin.wang@njit.edu}]{\inits{H.}\fnm{Haimin}~\lnm{Wang}}
\author[addressref={aff1,aff5}]{\inits{H.}\fnm{Haodi}~\lnm{Jiang}}
\author[addressref={aff1,aff3}]{\inits{Q.}\fnm{Qin}~\lnm{Li}}
\author[addressref={aff1,aff2}]{\inits{Y.}\fnm{Yasser}~\lnm{Abduallah}}
\author[addressref={aff1,aff3,aff4}]{\inits{Y.}\fnm{Yan}~\lnm{Xu}}
\address[id=aff1]{Institute for Space Weather Sciences, New Jersey Institute of Technology, University Heights, Newark, NJ 07102-1982, USA}
\address[id=aff2]{Department of Computer Science, New Jersey Institute of Technology, University Heights, Newark, NJ 07102-1982, USA}
\address[id=aff3]{Center for Solar-Terrestrial Research, New Jersey Institute of Technology, University Heights, Newark, NJ 07102-1982, USA}
\address[id=aff4]{Big Bear Solar Observatory, New Jersey Institute of Technology, 40386 North Shore Lane, Big Bear City, CA 92314-9672, USA}
\address[id=aff5]{Department of Computer Science, Sam Houston State University, Huntsville, TX 77341-2090, USA}

\runningauthor{C. Xu et al.}
\runningtitle{Super-Resolution of SOHO/MDI Magnetograms of ARs}

\begin{abstract}
Image super-resolution has been an important subject in image processing and recognition. 
Here, we present an attention-aided convolutional neural network (CNN) for solar image super-resolution. 
Our method, named SolarCNN, aims to enhance the quality of line-of-sight (LOS) magnetograms
of solar active regions (ARs) collected by the \textit{Michelson Doppler Imager} (MDI) 
on board the \textit{Solar and Heliospheric Observatory} (SOHO). 
The ground-truth labels used for training SolarCNN are the LOS magnetograms collected by the \textit{Helioseismic and Magnetic Imager} (HMI) on board the \textit{ Solar Dynamics Observatory} (SDO). 
Solar ARs consist of strong magnetic fields
in which magnetic energy can suddenly be released
to produce extreme space weather events, such as solar flares, coronal mass ejections, and solar energetic particles. 
SOHO/MDI covers Solar Cycle 23, which is stronger
with more eruptive events than Cycle 24. 
Enhanced SOHO/MDI magnetograms allow for better understanding and forecasting of violent events of space weather.
Experimental results show that SolarCNN improves the quality of SOHO/MDI magnetograms in terms of the structural similarity index measure (SSIM), Pearson's correlation coefficient (PCC), and the peak signal-to-noise ratio (PSNR).
\end{abstract}
\end{opening}

\section{Introduction}
\label{sec:intro}           
Deep learning, which is a subfield of machine learning, has drawn significant interest in recent years. 
It was originally used in speech recognition \citep{6639344}, 
natural language processing \citep{app11093986},
and computer vision \citep{DBLP:journals/access/HuTPW18}.
More recently, it has been applied to astronomy, astrophysics, and solar physics
\citep{2020ApJ...894...70L,2021ApJS..256...20J,2023SoPh..298....8E,2023SoPh..298....4M, 2023SoPh..298....6S}.
Here, we present a new 
deep-learning
method, specifically
an attention-aided convolutional neural network (CNN),
named SolarCNN, 
for solar image super-resolution. 
SolarCNN aims to enhance the quality of line-of-sight (LOS) magnetograms of solar active regions (ARs) 
collected by the \textit{Michelson Doppler Imager} \citep[MDI;][]{MDI} 
on board the \textit{Solar and Heliospheric Observatory} 
\citep[SOHO;][]{SOHO}. 
The ground-truth labels used for training SolarCNN
are the LOS magnetograms of the same ARs collected by 
the \textit{Helioseismic and Magnetic Imager} 
\citep[HMI;][]{HMI} on board 
the \textit{Solar Dynamics Observatory} 
\citep[SDO;][]{SDO}.
Training and test samples are collected from ARs in the HMI and MDI overlap period,
between 1 May 2010 and 11 April 2011. 

An AR on the solar disk usually consists of one or more sunspots and pores
that are formed because of the concentrations of strong magnetic fields. 
In the AR, magnetic energy can suddenly be released
to produce extreme space weather events, such as
solar flares \citep{1985SoPh...96..293M,
2011SSRv..158....5H, 2014ApJ...797...50A, Liu2019ApJ}, 
coronal mass ejections \citep{webb2012coronal, 2016SoPh..291.2017P,
Liu2020ApJ},
and solar energetic particles \citep{Abduallah2022ApJS,2022SoPh..297...32R}. 
SOHO/MDI covers Solar Cycle 23, which is stronger
with more eruptive events than Cycle 24. 
Enhanced SOHO/MDI magnetograms allow for better understanding and forecasting of violent events of space weather.
As indicated in \citet{2022ApJ...941...20L}, magnetograms with significantly reduced resolutions
would affect the accuracy of solar flare predictions. 

\begin{figure*}
    \centerline{
     \hspace{-0.075\textwidth}
     \includegraphics[width=1.00\columnwidth]{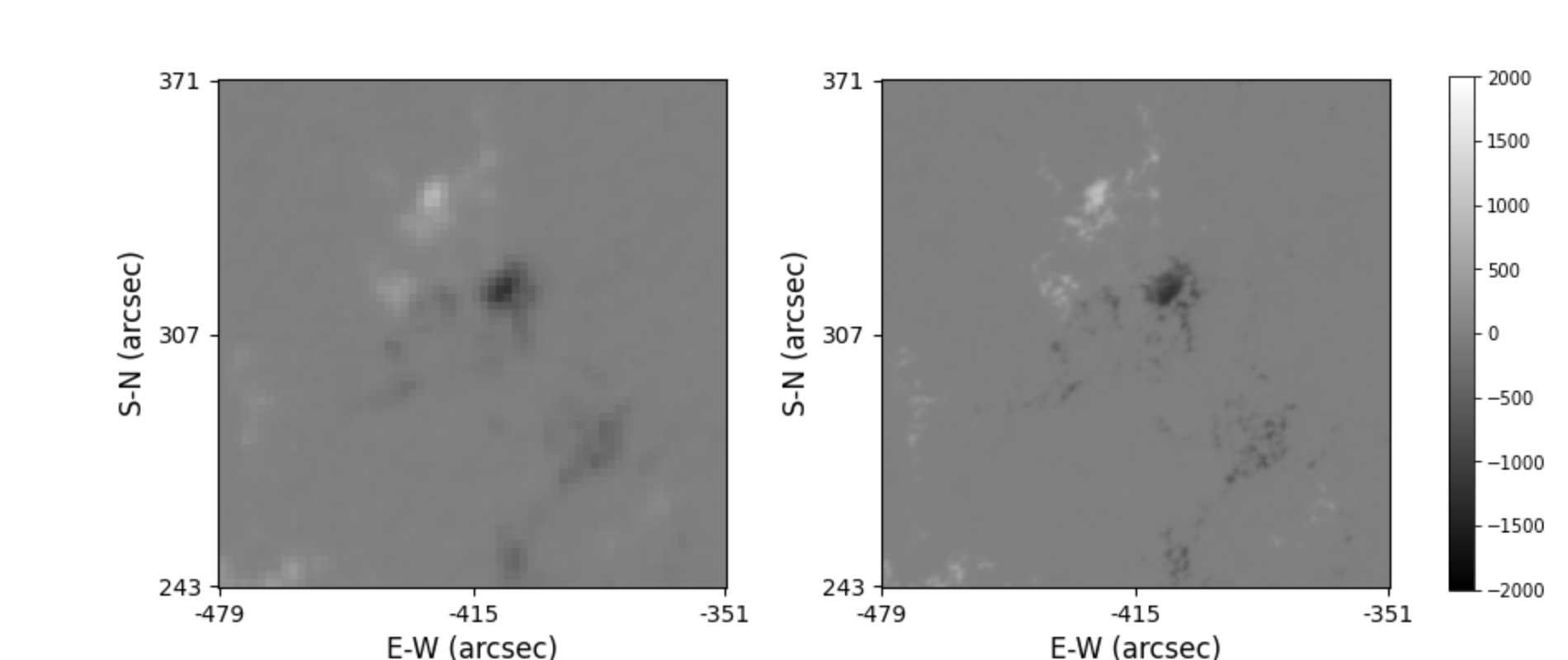}
    }
    \vspace{-0.115\textwidth}
    \centerline{
     \hspace{0.365 \columnwidth}  \color{white}{\textbf{a})} 
    \hspace{0.367 \columnwidth}  \color{white}{\textbf{b})} 
         \hfill}
    \vspace{+1.1cm}
	\caption{Comparison of LOS magnetograms, 
    taken by \textbf{a}) SOHO/MDI, and \textbf{b}) SDO/HMI. 
    Both images were taken from AR 11064 at 00:00:00 UT on 1 May 2010.}
    \label{fig:two_m}
\end{figure*}

SOHO/MDI and SDO/HMI were
designed to study the oscillations and magnetic fields
at the solar surface, or photosphere.
MDI was an older instrument, which was terminated in April 2011. 
HMI, the successor of MDI, is still on mission. 
Both MDI and HMI
provide full-disk LOS magnetograms,
where the LOS magnetograms of HMI have spatial and temporal resolutions higher
than those of MDI.
In addition, HMI provides vector magnetograms \citep{JLL-2023}.
Figure \ref{fig:two_m} compares the NOAA AR 11064 LOS magnetograms
collected by MDI (shown in Figure \ref{fig:two_m}a) and 
HMI (shown in Figure \ref{fig:two_m}b), 
respectively, at 00:00:00 UT on 1 May 2010.
It is evident from Figure \ref{fig:two_m} that the HMI magnetogram has a better spatial resolution
than the MDI magnetogram.

The remainder of this paper is organized as follows.
Section \ref{sec:rel} surveys related work.
Section \ref{sec:data} describes the data used in this study.
Section \ref{sec:meth} presents details of our SolarCNN model.
Section \ref{sec:E_R} reports experimental results, 
demonstrating that
SolarCNN performs well in terms of
commonly used image quality
assessment metrics \citep{sara2019image}, such as
the structural similarity index measure (SSIM), the
Pearson correlation coefficient (PCC), and the
peak signal-to-noise ratio (PSNR).
Section \ref{sec:con} presents a discussion and concludes the paper.

\section{Related Work} 
\label{sec:rel}
CNNs are very effective for
image processing and
have been extensively used by the communities of solar physics and space weather. 
They have been used to
infer line-of-sight velocities, Doppler widths, and vector magnetic fields from Stokes profiles \citep{2020ApJ...894...70L, 2022ApJ...939...66J}, 
track magnetic flux elements \citep{2020ApJS..250....5J}, 
trace H$\alpha$ fibrils \citep{2021ApJS..256...20J}, 
predict flares \citep{2018SoPh..293...48J}, 
detect filaments \citep{2019SoPh..294..117Z}, and so on. 
This motivates us to adopt CNNs for solar image super-resolution.
\citet{DBLP:journals/tmm/YangZTWXL19}
presented several 
deep-learning 
models for
single image super-resolution.
Other researchers performed image super-resolution using a multistage enhancement network \citep{DBLP:journals/access/HuangC22a}
or diffusion probabilistic models \citep{DBLP:journals/ijon/LiYCCFXLC22}. 
Nearly all models use convolution mechanisms to build multiple blocks and link these blocks with residuals. 
The residual structure was originally used to enhance the performance of the VGG network \citep{DBLP:conf/cvpr/HeZRS16}. 
It significantly improves 
deep-learning-based 
picture processing and has a large impact on image super-resolution methods \citep{8589454}. 
Inspired by the success of the residual structure, 
SolarCNN employs multiple blocks with convolutional layers
and residual links that link the multiple blocks.

To further improve the performance of SolarCNN, 
we include the Mish activation function. 
The Mish activation function is a modification of the ReLu activation function
and is used to obtain a loss lower than the ReLu activation function \citep{DBLP:conf/bmvc/Misra20}. 
The Mish activation function has been observed to perform better than
the ReLu activation function \citep{DBLP:journals/bspc/RahimHS21,DBLP:journals/ijon/LiYCCFXLC22}.
Furthermore, SolarCNN uses structures of a frequency channel attention network
(FcaNet) to increase its learning capacity \citep{DBLP:conf/iccv/QinZW021}.

In solar physics, several researchers have developed
deep-learning 
methods for super-resolution of solar images.
\citet{2018A&A...614A...5D} designed CNNs with
residual blocks to enhance HMI observations.
\citet{2020ApJ...897L..32R} and
\citet{2021ApJ...923...76D} 
proposed generative adversarial networks (GANs)
for super-resolution of HMI data.
\citet{2020ApJ...897L..32R} used a GAN model
to enhance HMI magnetograms,
validated by Hinode data.
\citet{2021ApJ...923...76D} employed a different GAN model to enhance HMI continuum images to the GST level.
\citet{2022ApJS..263...25S} used a diffusion probability model to enhance HMI continuum images.
In contrast to the above methods,
our work focuses on the enhancement of MDI data rather than HMI data.
Because of the difference in the data, 
SolarCNN's architecture significantly differs from the
architectures of the
previously developed 
deep-learning 
models for solar image super-resolution.

\section{Data}\label{sec:data}
We consider the overlap period of MDI and HMI,
between 1 May 2010 and 11 April 2011, 
in which both MDI and HMI data are available from the Joint Science Operations Center (JSOC) accessible at \url{jsoc.stanford.edu}.
Our dataset includes level 1.8 full-disk MDI magnetograms 
with \SI{2}{\arcsecond} per pixel
taken from the {\sf mdi.fd\_M\_96m\_lev182}
series of JSOC. 
HMI magnetograms with \SI{0.5}{\arcsecond} per pixel
from the {\sf hmi.M\_720s} series of JSOC 
are used as ground-truth labels. 
The cadence of the MDI magnetograms is 96 minutes, while the cadence of the HMI magnetograms is 12 minutes. 
The nearest HMI magnetogram is selected for each MDI image
to construct training pairs with an effective cadence of 96 min.
Our targets are ARs with unsigned flux peak 
$\geq$1500 G. 
The field of view (FOV) of each selected AR contains 256 $\times$ 256 pixels. 
The original size of an MDI image is 64 $\times$ 64.
To facilitate model training and evaluation, 
we extend the size to 256 $\times$ 256 through bilinear interpolation.
Thresholds of 
$\pm$ 2000 G 
are used, as this is
a typical range of AR magnetic fields
\citep{2020ApJ...897L..32R}. 
For the minority group of pixels with stronger field strengths, 
their values are set to 
$\pm$ 2000 G 
according to the polarities. 

Next, we normalize the magnetic field strengths in a magnetogram by dividing them by 2000, 
giving a range of [$-1$, 1].
ARs outside 
$\pm$ $70^{\circ}$ 
of the central meridian
are excluded to minimize projection effects.
This process yields a set of 1,569 pairs of aligned images (AR patches) 
from MDI and HMI. 
Among the 1,569 pairs, 
we select 1,493 pairs for model training and the remaining 76 pairs for model testing. 
The ARs from which the 1,493 training image pairs are taken differ from
the ARs from which the 76 test image pairs are collected. 
Thus, SolarCNN can make predictions on ARs that it has never seen during training. 
To avoid overfitting and
increase the diversity of the training set, 
image rotation is applied to the training samples, producing
5,972 pairs of aligned training images
and 76 pairs of aligned test images, where the size of each image is 256 $\times$ 256.
A random sample of 10\% of the training data is used for validation.

\section{Methodology}\label{sec:meth}
\begin{figure*}
\centering
\includegraphics[width=0.7\columnwidth]{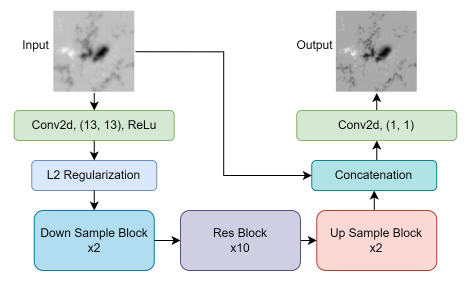}
\caption{Architecture of SolarCNN. After the initial 2D convolutional layer and regularization, 
the input image is downsampled by two consecutive Down Sample Blocks. Then, the data flow goes through ten Res Blocks to complete feature optimization. Next, the data flow passes through two Up Sample Blocks. Finally, the data flow is concatenated with the input image, where the concatenated result is sent to a 2D convolutional layer to obtain the output.}
\label{fig:mo_st}
\end{figure*}

SolarCNN employs downsampling and upsampling techniques, 
which are often used in 
deep-learning-based 
image super-resolution \citep{DBLP:journals/tmm/YangZTWXL19, DBLP:journals/access/HuangC22a, DBLP:journals/ijon/LiYCCFXLC22}. 
Figure \ref{fig:mo_st} shows the architecture of SolarCNN
for enhancing an MDI magnetogram to the HMI level.
The size of the input image is 256 × 256.  
The input image is first sent to a two-dimensional (2D) convolutional layer
with 64 filters (kernels) of size 13 $\times$ 13 followed by a ReLU activation function. 
The output of the 2D convolutional layer is sent to an L2 regularization layer, 
whose output is then sent to two Down Sample Blocks for downsampling, 
and ten Res Blocks for feature optimization. 
The output of the last Res Block is sent to two Up Sample Blocks. 

We also design an Fca Block as a subblock of the Res Block and Up Sample Block. 
The purpose of the Fca Block is to make the learned features self-update. 
Since SolarCNN aims to preserve and optimize as much as possible the original physical meaning of the input magnetogram, 
which is very different from the color images used in other color image super-resolution studies, 
we combine the input magnetogram and the output of the last Up Sample Block using a concatenation layer. 
The output of the concatenation layer is then sent to a 2D convolutional layer with 1 filter of size 1 $\times$ 1 to obtain the final output. 

The concatenation layer mentioned above is reminiscent of the connection structure used in U-Net \citep{Falk2019}.
The difference is that in U-Net, the connections, which exist between the encoder and the decoder, are for maintaining spatial information during the data resizing process, while
in SolarCNN, the concatenation layer, which exists outside the Down Sample Blocks and Up Sample Blocks, is
for generating an enhanced image based on the input magnetogram.

\begin{figure*}
\centerline{
\includegraphics[width=0.7\columnwidth]{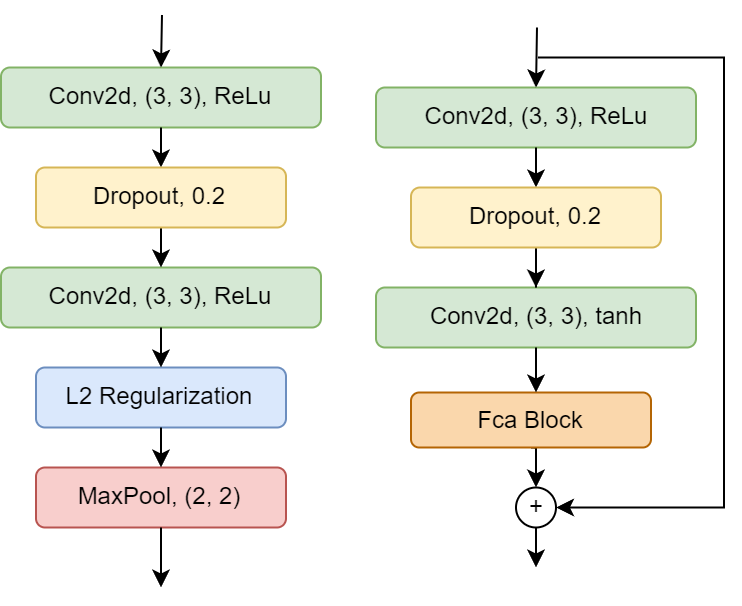}
}
\vspace{-0.55\textwidth}
\centerline{
\hspace{0.1 \columnwidth}  \color{black}{\textbf{a})} 
\hspace{0.3 \columnwidth}  \color{black}{\textbf{b})} 
\hfill}
\vspace{0.5\textwidth}
\centerline{
\includegraphics[width=0.7\columnwidth]{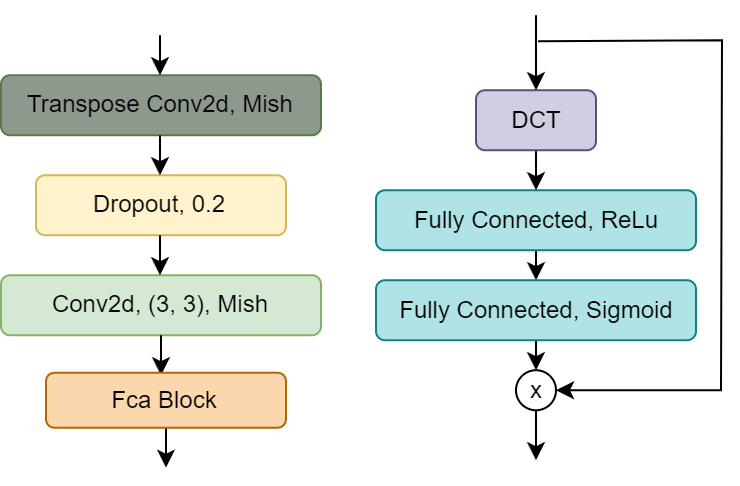}
}
\vspace{-0.45\textwidth}
\centerline{
\hspace{0.1 \columnwidth}  \color{black}{\textbf{c})} 
\hspace{0.3 \columnwidth}  \color{black}{\textbf{d})} 
\hfill}
\vspace{0.4\textwidth}
\caption{Configuration details of the 
\textbf{a}) Down Sample Block, 
\textbf{b}) Res Block, 
\textbf{c}) Up Sample Block, and 
\textbf{d}) Fca Block. 
In the Down Sample Block, the data flow first passes through two 2D convolutional layers, followed by L2 regularization and max pooling. A dropout rate of 0.2 is added between the two 2D convolutional layers. In the Res Block, the data flow passes through two 2D convolutional layers with a dropout rate of 0.2 between them. The data flow then passes through an Fca Block. Finally, the output data is residual connected with the input flow to obtain the output of the Res Block. In the Up Sample Block, the data flow passes through a transposed 2D convolutional layer and
another 2D convolutional layer with a dropout rate of 0.2 between them. Then, the data flow goes through an Fca Block to obtain the output of the Up Sample Block. In the Fca Block, the data flow goes through a discrete cosine transform (DCT) layer, then passes through two fully connected layers, and finally is multiplied by the input of the Fca Block to obtain the output
of the Fca Block. 
The Res Block and the Up Sample Block
contain an Fca Block as a subblock.}
\label{fig:b12_st}
\end{figure*}

Figure \ref{fig:b12_st} presents the configuration details of the Down Sample Block, Res Block, Up Sample Block, and Fca Block. 
In the Down Sample Block shown in Figure \ref{fig:b12_st}a, 
the input image is first convolved twice and regularized once with L2. 
Then it goes through a max-pooling layer for downsampling. 
The first 2D convolutional layer has 64 filters of size 3 $\times$ 3, 
and the second 2D convolutional layer has 128 filters of size 3 $\times$ 3. 
These filters (kernels) are followed by a ReLu activation function. 
In addition, we add a dropout layer between the two 2D convolutional layers. 
The dropout rate is 0.2. The kernel size of the max-pooling layer is 2 $\times$ 2. 
After that, the number of filters doubles for each new Down Sample Block connected in the model. 
We use two Down Sample Blocks. 
Thus, the input size of the Down Sample Blocks is
256 $\times$ 256 $\times$ 64, 
and the output size is 64 $\times$ 64 $\times$ 256.

In the Res Block shown in Figure \ref{fig:b12_st}b, 
the input data flow passes through two 2D convolutional layers. 
The activation function of the first 2D convolutional layer is ReLu, 
and the activation function of the second 2D convolutional layer is tanh. 
Both 2D convolutional layers have 256 filters of size 3 $\times$ 3. 
We also add a dropout layer between the two 2D convolutional layers
with a dropout rate of 0.2. 
The output data flow of the second 2D convolutional layer passes through an Fca Block. 
We use ten Res Blocks in SolarCNN. The input and output size of this part is 64 $\times$ 64 $\times$ 256.

In the Up Sample Block shown in Figure \ref{fig:b12_st}c, 
the input flow passes through a transposed 2D convolutional layer, a dropout layer, 
and another 2D convolutional layer (the second 2D convolutional layer) to obtain the output flow. 
The transposed 2D convolutional layer has 128 filters of size 2 $\times$ 2 with a stride of (2, 2), 
followed by a Mish activation function. 
The dropout rate of the dropout layer is 0.2. 
The second 2D convolutional layer has 128 filters of size 3 $\times$ 3, also followed by a Mish activation function. 
The data flow of the second 2D convolutional layer passes through an Fca Block. 
For each new Up Sample Block connected in the model, the number of filers is halved. 
The number of Up Sample Blocks is the same as the number of Down Sample Blocks. 
Therefore, the input size of the Up Sample Blocks is 64 $\times$ 64 $\times$ 256, and the output size is 256 $\times$ 256 $\times$ 64.

In the Fca Block shown in Figure \ref{fig:b12_st}d, 
the input features first go through a discrete cosine transform (DCT) layer and then go through two fully connected layers. 
In the first fully connected layer, the number of nodes is twice the number of filters for the input features, and the activation function is ReLu.
In the second fully connected layer, the number of nodes is the same as the number of filters for the input features, and the activation function is Sigmoid.
The result of the second fully connected layer is filled to the same shape as the input features. 
The product of the input features and the filled result is the output of the Fca Block. 

Table \ref{tab:par} summarizes the configuration details of SolarCNN.
Table \ref{tab:hyp} presents the settings of SolarCNN's hyperparameters. 
We use a mixture of SSIM (to be defined in Section \ref{sec:E_R})
and L1 loss as the loss function for the SolarCNN model. 
SSIM is a positive direction indicator, 
so our loss function is the sum of
the L1 loss and the negative of SSIM. 
The optimizer we choose is Adam with a learning rate of 1 $\times$ $10^{-5}$  
and a weight decay rate of 1 $\times$ $10^{-5}$.

\begin{table}
\renewcommand{\arraystretch}{1.}
\caption{Configuration details of SolarCNN.}
\label{tab:par}
{
\centering
\begin{tabular}{c c c c}
\hline
\bfseries Block & \bfseries Layer & \bfseries Kernel Size & \bfseries Activation\\
\hline		
& Input Convolutional Layer & 13 × 13 & ReLu\\
\hline
Down Sample Block & Convolutional Layer 1 & 3 × 3 & ReLu\\
& Convolutional Layer 2 & 3 × 3 & ReLu\\
& Max Pool Layer & 2 × 2 &\\
\hline
Res Block & Convolutional Layer 1 & 3 × 3 & ReLu\\
& Convolutional Layer 2 & 3 × 3 & tanh\\
\hline
Up Sample Block & Transposed Convolutional Layer & 2 × 2 & Mish\\
& Convolutional Layer & 3 × 3 & Mish\\
\hline
Fca Block & Fully Connected Layer 1 & & ReLu\\
& Fully Connected Layer 2 & & Sigmoid\\
\hline
& Output Convolutional Layer & 1 × 1 & \\
\hline
\end{tabular}
}
\end{table}	

\begin{table}
\renewcommand{\arraystretch}{1.}
\caption{The training hyperparameters of SolarCNN.}
\label{tab:hyp}
{
\centering
\begin{tabular}{c c c c c}
\hline
\bfseries Optimizer &\bfseries Learning Rate &\bfseries Weight Decay &\bfseries Batch Size &\bfseries Epoch \\
\hline
Adam & 1 $\times$ $10^{-5}$ & 1 $\times$ $10^{-5}$ & 64 & 1000 \\
\hline
\end{tabular}
}
\end{table}	

Although we use image rotation in the image augmentation process to increase the data complexity, 
there is still the possibility of overfitting. 
During training, the loss on the training set decreases steadily, whereas the loss on the validation set may bounce back after decreasing to a certain level. 
To avoid overfitting, 
we adopt an early stop approach, which works as follows.  
When the loss in the validation set can no longer drop and starts to rebound slightly, the SolarCNN model training is terminated and the model parameters are saved.

\section{Experiments and Results}\label{sec:E_R}
\subsection{Evaluation Metrics}
\label{sec:E_R:isr}
We conduct a series of experiments to assess the components of SolarCNN
and compare it with related methods.
The evaluation metrics used in the experiments include
the structural similarity index measure (SSIM), Pearson's correlation coefficient (PCC), and
peak signal-to-noise ratio (PSNR). 

The SSIM of two equal-sized magnetograms $A$ and $B$, each having $n$ pixels, is defined as:
\begin{equation}
    \mbox{SSIM} = \frac{(2\mu_{A}\mu_{B}+C_{1})(2\sigma_{A,B}+C_{2})}{({\mu_{A}^{2}}+{\mu_{B}^{2}}+C_{1})({\sigma_{A}^{2}}+{\sigma_{B}^{2}}+C_{2})},
\end{equation}
where 
$\mu_{A}$ is the mean of the
pixel values
of all pixels in $A$, 
$\mu_{B}$ is the mean of the 
pixel values
of all pixels in $B$,
$\sigma_{A,B}$ is the covariance of $A$ and $B$, 
$\sigma_{A}^{2}$ is the variance of the 
pixel values
of all pixels in $A$,
$\sigma_{B}^{2}$ is the variance of the 
pixel values
of all pixels in $B$,
$C_{1}$ and $C_{2}$ are constants.
The covariance of $A$ and $B$ is:
\begin{equation}
 \sigma_{A,B} = \frac{1}{n} \sum_{i=1}^{n} 
 (A_{i} - \mu_{A})(B_{i} - \mu_{B}),
\end{equation}
where $A_{i}$ ($B_{i}$, respectively) is the 
pixel value
of the $i$th pixel in $A$ ($B$, respectively).
SSIM ranges from $-1$ to 1, with a larger SSIM value indicating a greater similarity between $A$ and $B$. 

PCC represents the strength of correlation between $A$ and $B$,
which is defined as:
\begin{equation}
    \mbox{PCC} = \frac{\sum_{i=1}^{n}(A_i-\mu_{A})(B_i-\mu_{B})}{\sqrt{\sum_{i=1}^{n}{(A_i-\mu_{A})^2}} 
    \sqrt{\sum_{i=1}^{n}{(B_i-\mu_{B})^2}}}.
\end{equation}
PCC also ranges from $-1$ to $1$,
with a higher value indicating a stronger correlation
between $A$ and $B$. 
A scatter plot is often used to visualize
the correlation between $A$ and $B$.
The closer the scatter plot dots to the diagonal line $Y = X$,
the stronger the correlation between $A$ and $B$.

PSNR is commonly used in image super-resolution research,
which is defined as:
\begin{equation}
\mbox{PSNR} = 10 \log_{10} (\frac{\mbox{MAX}^2}{\mbox{MSE}}),
\end{equation}
where MAX is the maximum fluctuation of 
pixel values
in a magnetogram.
Because we limit the maximum magnetic field strength of a pixel to 2000 G, 
MAX is set to 4000 to take into account both positive and negative magnetic flux regions.
The mean squared error (MSE) is: 
\begin{equation}
    \mbox{MSE} = \frac{1}{n}\sum_{i=1}^{n}{(A_i-B_i)}^{2}.
\end{equation}
The larger the PSNR, the closer $A$ to $B$. 

\subsection{Ablation Study}
\label{sec:E_R:cv}
In conducting ablation tests to assess the components of SolarCNN,
we considered seven
variants of SolarCNN:
SolarCNN-L,
SolarCNN-R,
SolarCNN-F,
SolarCNN-LR,
SolarCNN-LF,
SolarCNN-RF, and
SolarCNN-LRF.
SolarCNN-L denotes SolarCNN with the L2 regularization layer being removed, where
the L2 regularization layer is used to prevent overfitting.
SolarCNN-R denotes SolarCNN with the Res Blocks being removed.
SolarCNN-F denotes SolarCNN with the Fca Blocks being removed
where the discrete cosine transform (DCT) layer in an Fca Block
is used for contrast enhancement.
SolarCNN-LR denotes SolarCNN with the
L2 regularization layer and Res Blocks being removed.
SolarCNN-LF denotes SolarCNN with the
L2 regularization layer and Fca Blocks being removed.
SolarCNN-RF denotes SolarCNN with the
Res and Fca Blocks being removed.
SolarCNN-LRF denotes SolarCNN with the
L2 regularization layer, Res and Fca Blocks being removed.
Table \ref{tab:abl_hmi} presents the metric values
of the eight models. 
 
\begin{table}
\renewcommand{\arraystretch}{1.}
\caption{Results of the ablation study.}
\label{tab:abl_hmi}
{
\centering
    \begin{tabular}{l|c|c|c}
    \hline	
    \bfseries Model & \bfseries SSIM & \bfseries PCC & \bfseries PSNR \\
    \hline
    SolarCNN & \bfseries 0.9039 & \bfseries 0.8842& \bfseries 37.40 \\
    SolarCNN-L & 0.8884 & 0.8703 & 36.23 \\
    SolarCNN-R & 0.8857 & 0.8765 & 35.89 \\
    SolarCNN-F & 0.9017 & 0.8818 & 37.34 \\
    SolarCNN-LR & 0.8702 & 0.8683 & 35.31 \\
    SolarCNN-LF & 0.8784 & 0.8697 & 36.07 \\
    SolarCNN-RF & 0.8687 & 0.8701 & 35.34 \\
    SolarCNN-LRF & 0.8622 & 0.8664 & 35.13 \\
    \hline
      None & 0.8334 &  0.8627 &  34.73 \\
    \hline
    \end{tabular}    }
    \vspace{-0.05\textwidth}
\end{table}

The eight models use the same training and test sets
as described in Section \ref{sec:data},
and the same hyperparameter values as shown in Table \ref{tab:hyp}.
The results in Table \ref{tab:abl_hmi}
are the averages of the metric values for all the test magnetograms in the test set.
For each test magnetogram, we
compare its enhanced image with 
the corresponding ground-truth label
(HMI magnetogram)
and use the formulas in Section \ref{sec:E_R:isr}
to calculate the metric values.
For comparison purposes, we also include the case where no model is used,
denoted by ``None'', in Table \ref{tab:abl_hmi}.
The metric values in the ``None'' row in Table \ref{tab:abl_hmi}
are obtained by comparing each MDI magnetogram directly with
the corresponding HMI magnetogram
without using any model.
The best metric values in the table
are in bold.

It can be seen from Table \ref{tab:abl_hmi} that
SolarCNN performs best with the largest value in each metric.
These results
indicate the importance and usefulness of the
L2 regularization layer,
Res and Fca Blocks to improve the performance of the proposed method.
SolarCNN achieves, on average, an SSIM of 0.9039,
PCC of 0.8842,
and PSNR of 37.40
when comparing
the SolarCNN-enhanced MDI magnetograms with their corresponding ground-truth labels
(HMI magnetograms). 
When comparing the original MDI magnetograms
with their corresponding HMI magnetograms,
we obtain, on average, an SSIM of 0.8334,
PCC of 0.8627, 
and PSNR of 34.73.
These numbers indicate that the SolarCNN-enhanced MDI magnetograms are closer to
the corresponding HMI magnetograms than
the original MDI magnetograms,
demonstrating the effectiveness of the proposed method.

\subsection{Comparative Study}
In this experiment, we compare SolarCNN with two related methods.
The first method, named CNNr, is also a CNN model with
residual blocks.
The CNNr architecture is inspired by the work of
\citet{2018A&A...614A...5D} for the enhancement of the SDO/HMI image.
The residual blocks and the up sample blocks of CNNr,
together with the optimizer and loss function,
are taken from \citet{2018A&A...614A...5D}.
To handle the image sizes at hand,
we added down-sample blocks of SolarCNN to CNNr.
The other hyperparameters of CNNr are the same as those of SolarCNN.
The second is the bicubic method \citep{2020ApJ...897L..32R},
which uses mathematical interpolation to enhance the image.

\begin{figure*}
\centering{
\includegraphics[width=0.7\columnwidth]{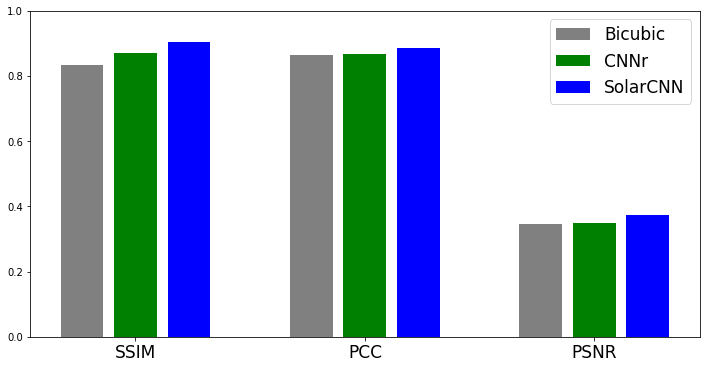}
}
\caption{Performance comparison between SolarCNN and two related methods (CNNr and the bicubic method).
}
\label{fig:math_hmi}
\end{figure*}

Figure \ref{fig:math_hmi} compares the three methods,
where the results are the averages of the metric values for
all the test magnetograms in the test set.
For each test MDI magnetogram, we
compare its enhanced image with the corresponding ground-truth label
(HMI magnetogram)
and calculate the metric values.
To facilitate visualization, the PSNR values
in Figure \ref{fig:math_hmi} are obtained by dividing the original PSNR values by 100.
It can be seen from Figure \ref{fig:math_hmi} that
SolarCNN is the best of the three methods.
The two CNN models (SolarCNN and CNNr) perform better than the bicubic method. 

\subsection{Case Studies}
\begin{figure*}
    \centerline{
    \hspace{-0.023 \textwidth}
     \includegraphics[width=1.2\columnwidth]{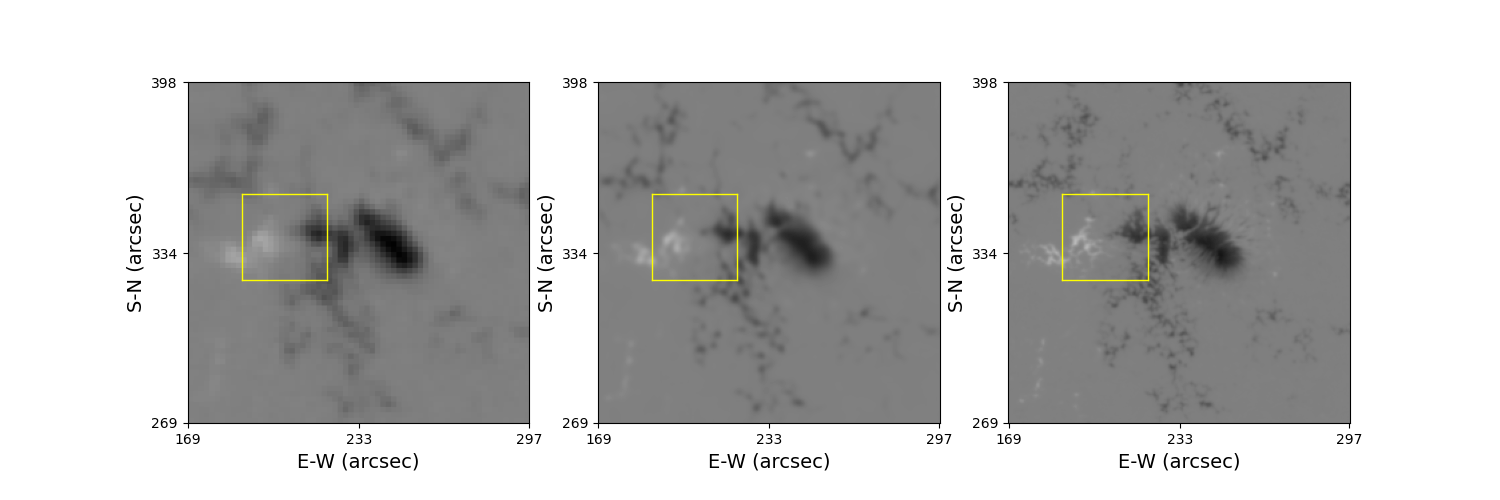}
    }
    \vspace{-0.11\textwidth}
    \centerline{
      \hspace{0.261\textwidth}  \color{white}{\textbf{a})}
      \hspace{0.281\textwidth}  \color{white}{\textbf{b})}
      \hspace{0.281\textwidth}  \color{white}{\textbf{c})}
         \hfill}
    \vspace{-0.015\textwidth}
    \centerline{
    \hspace{-0.023 \textwidth}
     \includegraphics[width=1.2\columnwidth]{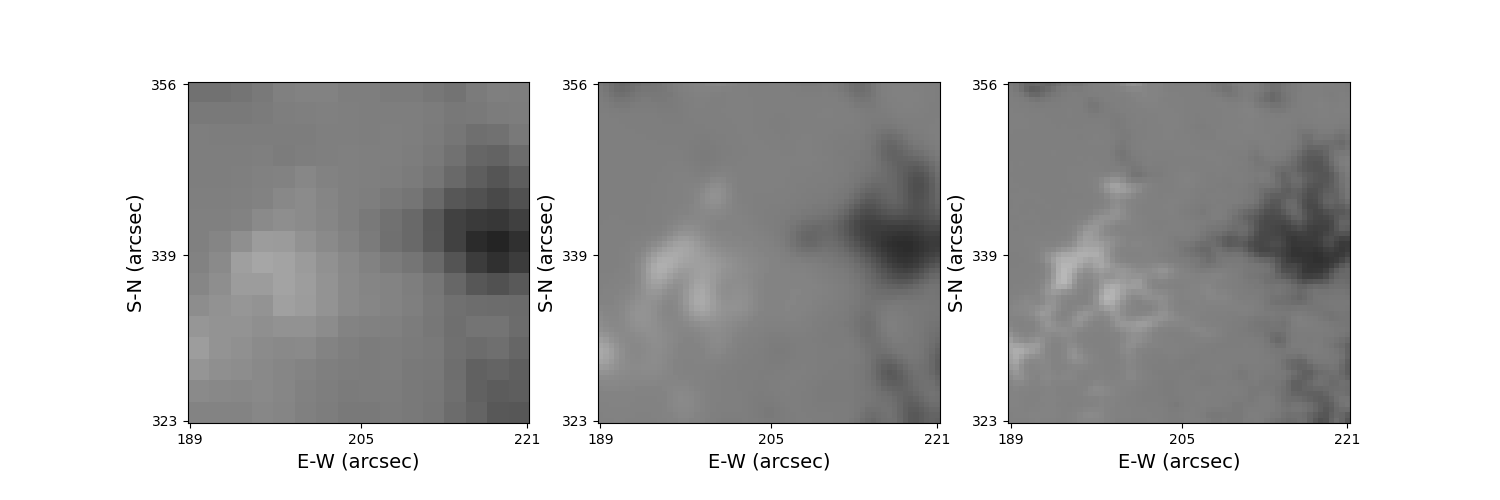}
    }
    \vspace{-0.11\textwidth}
    \centerline{
      \hspace{0.261 \textwidth}  \color{white}{\textbf{d})}
      \hspace{0.281\textwidth}  \color{white}{\textbf{e})}
      \hspace{0.281\textwidth}  \color{white}{\textbf{f})}
         \hfill}
    \vspace{0.04\textwidth}
    \centerline{
    \hspace{-0.045\textwidth}
    \includegraphics[width=1\columnwidth]{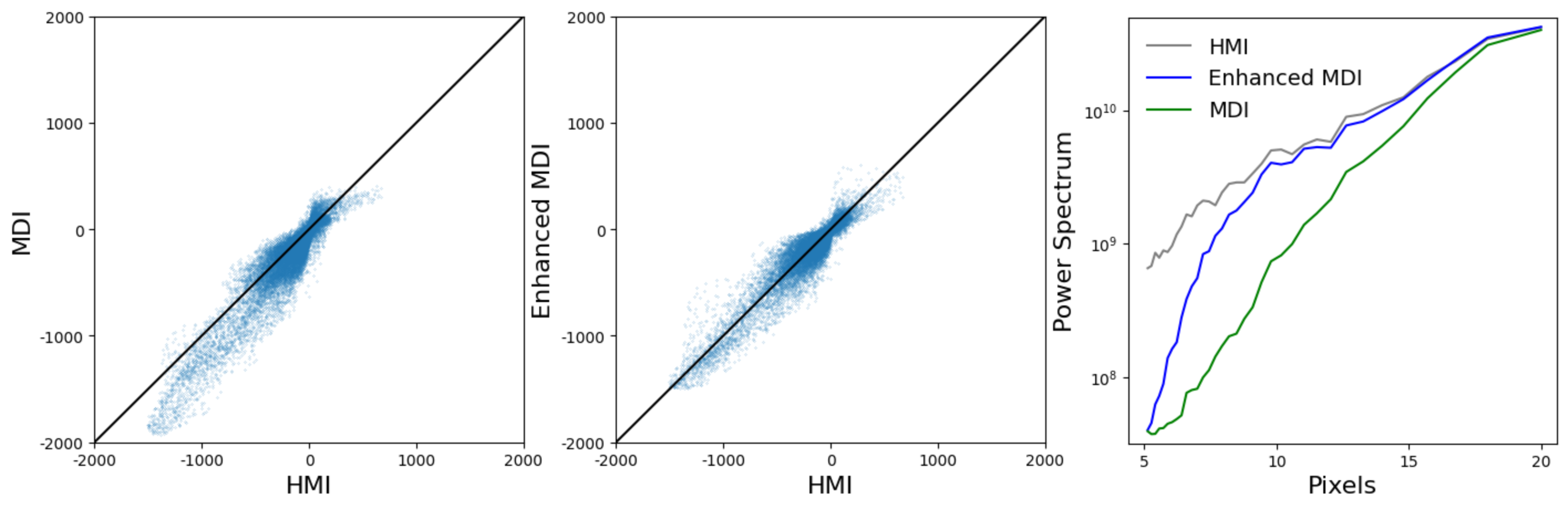}
    }
    \vspace{-0.089\textwidth}
    \centerline{
      \hspace{0.262 \textwidth}  \color{black}{\textbf{g})}
      \hspace{0.281\textwidth}  \color{black}{\textbf{h})}
      \hspace{0.283\textwidth}  \color{black}{\textbf{i})}
         \hfill}
    \vspace{0.05\textwidth}
	\caption{Comparison among an MDI magnetogram,
   its SolarCNN-enhanced magnetogram, and the corresponding
   HMI magnetogram of AR 11183 at 
   20:48:00 UT
   on 2 April 2011.
   \textbf{a}) MDI magnetogram. 
   \textbf{b}) Enhanced MDI magnetogram.
   \textbf{c}) HMI magnetogram.
   \textbf{d}) FOV of the region highlighted by the yellow box in \textbf{a}).
    \textbf{e}) FOV of the region highlighted by the yellow box in 
   \textbf{b}).
    \textbf{f}) FOV of the region highlighted by the yellow box in \textbf{c}).
   \textbf{g}) Scatter plot of the MDI magnetogram versus the HMI magnetogram.
   \textbf{h}) Scatter plot of the enhanced MDI magnetogram versus the HMI magnetogram.
   \textbf{i}) Azimuthally averaged power spectrum of the three magnetograms.}\label{fig:vis}
\end{figure*}

Here, we present several case studies to further demonstrate the effectiveness of SolarCNN.
Figure \ref{fig:vis} shows an MDI magnetogram, its enhanced magnetogram, and
the corresponding HMI magnetogram of
AR 11183 at
20:48:00 UT
on 2 April 2011. 
The top row in the figure displays,
from left to right,
the MDI magnetogram, the
enhanced MDI magnetogram, and the
HMI magnetogram.
The second row displays,
from left to right,
the FOV (field of view) of the region highlighted by the yellow box
in the corresponding magnetogram in the top row.
Visually, the enhanced MDI magnetogram is closer to the HMI magnetogram
than the original MDI magnetogram.
Quantitatively,
we obtain an SSIM of 0.8984,
PCC of 0.8897,
and PSNR of 37.04
when comparing the enhanced MDI magnetogram with the HMI magnetogram, and we obtain an SSIM of 0.8240,
PCC of 0.8691,
and PSNR of 33.23
when comparing the original MDI magnetogram with the HMI magnetogram.

Figure \ref{fig:vis}g 
shows the scatter plot
between the MDI magnetogram and the HMI magnetogram.
Figure \ref{fig:vis}h shows the scatter plot
between the enhanced MDI magnetogram and the HMI magnetogram. 
Clearly, the enhanced MDI magnetogram correlates better with the HMI magnetogram than the original MDI magnetogram.
We note that the enhanced MDI magnetogram is close to the HMI magnetogram
in not only resolution but also magnetic field strengths.

Figure \ref{fig:vis}i shows the azimuthally averaged power spectrum of the three magnetograms,
which is a measure of the amount of information in an image based on the Fourier transform \citep{2009A&A...503..225W, 2021ApJ...923...76D}.
It can be seen from Figure \ref{fig:vis}i
that the HMI magnetogram has the most information
with the highest resolution, 
and the enhanced MDI magnetogram is closer to the HMI magnetogram in terms of information and resolution than the MDI magnetogram.

\begin{figure*}
    \centerline{
    \hspace{-0.023 \textwidth}
     \includegraphics[width=1.\columnwidth]{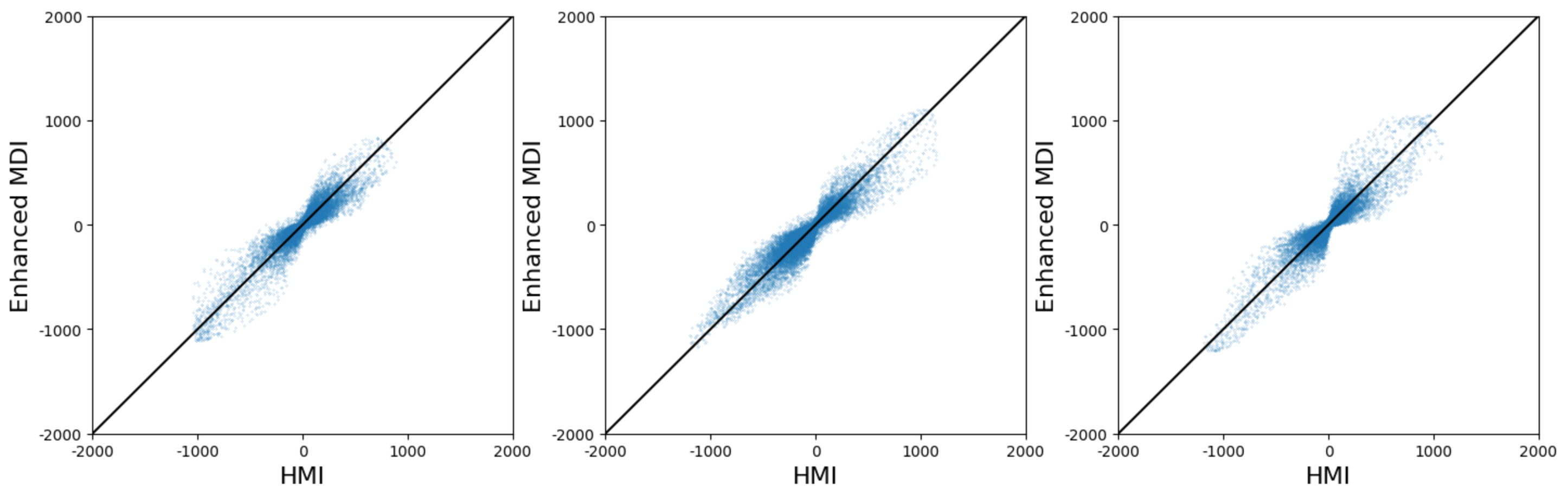}
    }
    \vspace{-0.09\textwidth}
    \centerline{
      \hspace{0.261\textwidth}  \color{black}{\textbf{a})}
      \hspace{0.281\textwidth}  \color{black}{\textbf{b})}
      \hspace{0.281\textwidth}  \color{black}{\textbf{c})}
         \hfill}
    \vspace{0.05\textwidth}
    \caption{Scatter plots of the SolarCNN-enhanced MDI magnetograms 
    versus the corresponding HMI magnetograms from \textbf{a}) AR 11185 at 12:48:00 UT on 6 April 2011, 
   \textbf{b}) AR 11186 at 16:00:00 UT on 9 April 2011, and \textbf{c}) AR 11189 at 00:00:00 UT on 9 April 2011, respectively.}
   \label{fig:h5a}
\end{figure*}

Figure \ref{fig:h5a} presents scatter plots of the enhanced MDI magnetograms versus the corresponding HMI magnetograms
from three additional active regions (ARs).
These active regions are AR 11185 at 12:48:00 UT on 6 April 2011,  
AR 11186 at 16:00:00 UT on 9 April 2011, and
AR 11189 at 00:00:00 UT on 9 April 2011.
It can be seen from 
Figure \ref{fig:h5a}
that SolarCNN transforms the MDI magnetograms into HMI-like magnetograms
in the sense that the enhanced MDI magnetograms are close to the
corresponding HMI magnetograms
in magnetic field strengths, a finding consistent with that in Figure \ref{fig:vis}h. 
When considering all 76 samples in the test set,
the factor between the magnetic field strengths of the enhanced MDI magnetograms and 
those of the corresponding HMI magnetograms
is approximately 0.937,
showing that SolarCNN can generally transform MDI magnetograms into HMI-like 
magnetograms.

\subsection{Comparison with Hinode Data}
To further understand the behavior of SolarCNN,
we applied it to an active region outside
the overlap period of MDI and HMI (between 1 May 2010 and
11 April 2011).
Specifically, we picked an MDI magnetogram of AR 11024 on 7 July 2009, and
used it as input to the trained SolarCNN model.
We compared the enhanced MDI magnetogram with the corresponding magnetogram 
of the same AR at the same day obtained from Hinode/SP. 
Hinode/SP (Spectro-Polarimeter) is an instrument designed to
analyze the Sun's magnetic fields and their influence on solar activity \citep{2008SoPh..249..167T}.
The resolution of Hinode/SP is \SI{0.16}{\arcsecond} per pixel.

\begin{figure*}
	\centering
 \includegraphics[width=0.9\columnwidth]{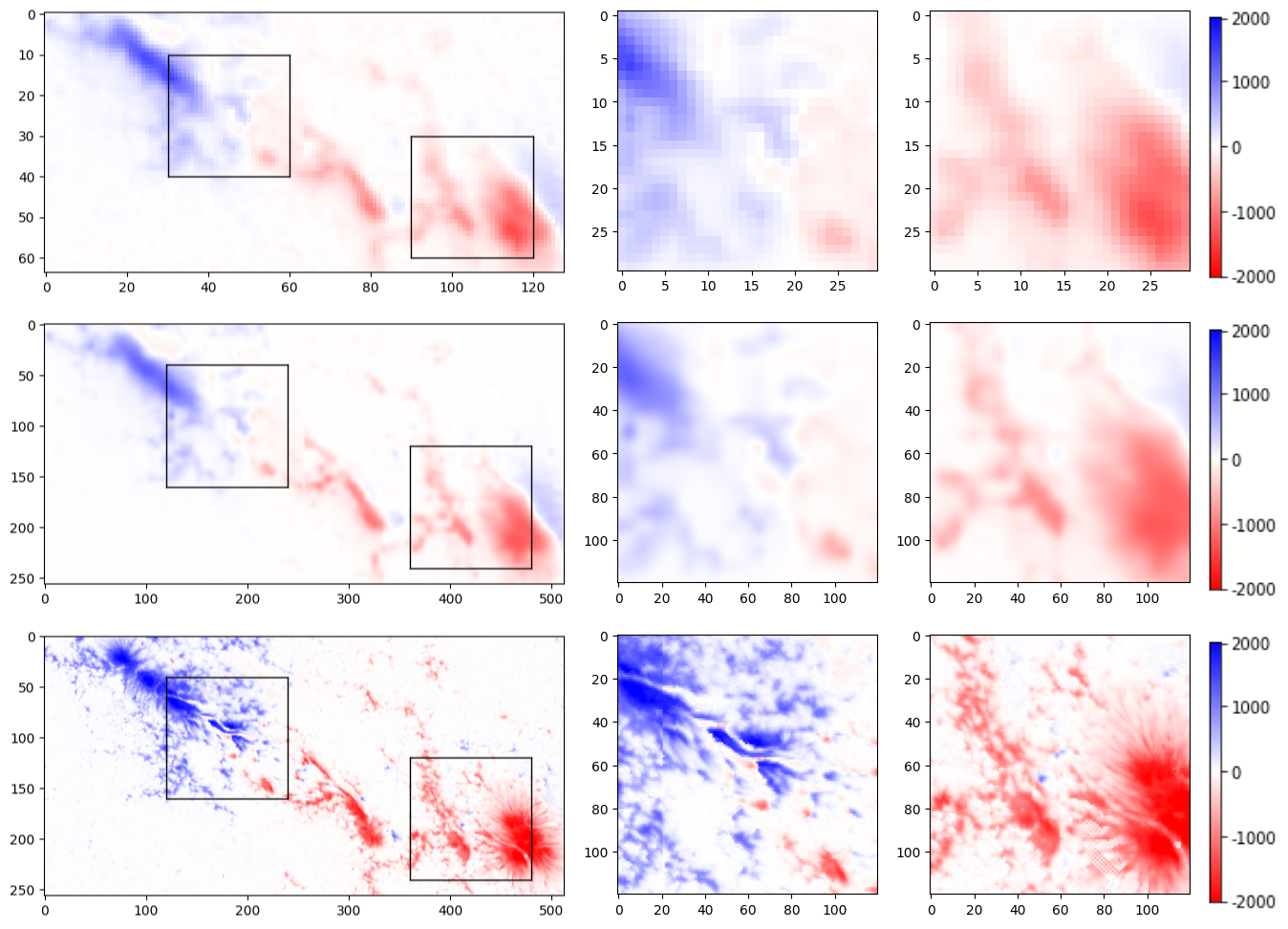}
    \vspace{+0.1cm}
	\caption{Comparison among an MDI magnetogram (top),
   its SolarCNN-enhanced magnetogram (middle), and the corresponding Hinode/SP magnetogram (bottom) of AR 11024 on 7 July 2009.}\label{fig:hin}
\end{figure*}

Figure \ref{fig:hin} presents the comparison result. 
The first row shows the MDI magnetogram and the FOVs
of the two regions highlighted by the black boxes in the MDI magnetogram.
The second row shows the SolarCNN-enhanced MDI magnetogram
and the FOVs of the two regions highlighted by the black boxes in the enhanced MDI magnetogram.
The third row shows the Hinode/SP magnetogram
and the FOVs
of the two regions highlighted by the black boxes 
in the Hinode/SP magnetogram.
HMI did not exist in 2009, so no HMI image is shown in Figure \ref{fig:hin}.

It can be seen from Figure \ref{fig:hin} that
the SolarCNN-enhanced MDI magnetogram has a higher resolution than the
MDI magnetogram, although the Hinode/SP magnetogram
is the most clear.
This is understandable given that SolarCNN is trained by HMI magnetograms
and its output is, at best, equivalent to an HMI magnetogram.
On the other hand, the resolution of a Hinode/SP magnetogram is approximately three times that of an HMI magnetogram, and therefore
the Hinode/SP magnetogram is more clear than
the SolarCNN-enhanced MDI magnetogram in Figure \ref{fig:hin}.

\section{Discussion and Conclusion}\label{sec:con}
In this paper, we present a new 
deep-learning 
method (SolarCNN) for
image super-resolution of solar ARs.
We use SolarCNN to enhance MDI magnetograms to the HMI level.
Training, test, and validation data are taken from the
HMI and MDI overlap period,
between 1 May 2010 and 11 April 2011,
when MDI and HMI obtained data simultaneously.
Our experimental results show that the proposed method
works well in terms of SSIM, PCC, and PSNR.
Furthermore, our ablation study indicates that the
L2 regularization,
residual structures, and
Fca mechanism
used in the SolarCNN model are effective,
improving the performance of the model.
Our comparative study shows that SolarCNN
performs better than
two related methods.

An MDI magnetogram is approximately
1.4 times stronger than an HMI magnetogram in terms of magnetic field strengths 
\citep{2012SoPh..279..295L},
though this factor is close to 1.1 
when examining all the 76 samples in our test set.
The decrease in the factor is probably due to the data reprocessing at JSOC.
On the other hand, 
the factor between the magnetic field strengths of 
the SolarCNN-enhanced MDI magnetograms 
and those of the corresponding HMI magnetograms
in our test set
is approximately 0.937.
We divide the magnetic field strength of each pixel of a test MDI magnetogram by 1.1, and 
calculate the average of evaluation metric values between the resulting
MDI magnetograms and corresponding HMI magnetograms in the test set.
We obtain an SSIM of 0.8480, PCC of 0.8702, and PSNR of 35.80.
In contrast, SolarCNN achieves an SSIM of 0.9039, PCC of 0.8842, and PSNR of 37.40 (see Table \ref{tab:abl_hmi}).
This result shows that SolarCNN not only increases the resolution of MDI
magnetograms but also changes the pixel values of MDI magnetograms,
transforming MDI magnetograms into HMI-like magnetograms.

As indicated above, 
MDI and HMI observations overlap for approximately 1 year. 
Consequently, we have a relatively small dataset. 
To further understand the performance and generalization of SolarCNN, 
we conducted an additional experiment
based on a cross-validation scheme. 
This is a standard technique for assessing
how well a 
machine-learning 
model generalizes to new, unseen data.
Specifically, our dataset covers 12 months in total.
We divided the dataset into 12 different subsets by month.
In the run $i$, we used the subset $i$ as test data and
the union of the other 11 subsets as training data.
There are 12 subsets and, therefore, 12 runs.
The mean values of SSIM, PCC and PSNR
obtained by SolarCNN
over the 12 runs
are 0.9171, 0.8911, 38.31 respectively.
These results are similar to the metric values obtained by SolarCNN
in Table \ref{tab:abl_hmi}.

On the basis of the above results, we conclude that
SolarCNN is a feasible tool for enhancing
the quality of SOHO/MDI magnetograms of solar ARs using SDO/HMI data.

\begin{acks}[Acknowledgments]
The authors thank the handling
editor and anonymous referee for constructive comments and suggestions. 
We also thank members of the Institute
for Space Weather Sciences for fruitful discussions.
SOHO is an international cooperation project between ESA and NASA. 
SDO is a NASA mission.  
The SolarCNN model is implemented in Python and TensorFlow. 
\end{acks}

\begin{fundinginformation}
This work was supported in part by U.S. NSF grants 
AGS-1927578, AGS-2149748, AGS-2228996, and OAC-2320147.
\end{fundinginformation}

\bibliographystyle{spr-mp-sola}

\begin{thebibliography}{44}
\ifx\bisbn     \undefined \def\bisbn  #1{ISBN #1}\fi
\ifx\binits    \undefined \def\binits#1{#1}\fi
\ifx\bauthor   \undefined \def\bauthor#1{#1}\fi
\ifx\batitle   \undefined \def\batitle#1{#1}\fi
\ifx\bjtitle   \undefined \def\bjtitle#1{\textit{#1}}\fi
\ifx\bvolume   \undefined \def\bvolume#1{\textbf{#1}}\fi
\ifx\byear     \undefined \def\byear#1{#1}\fi
\ifx\bissue    \undefined \def\bissue#1{#1}\fi
\ifx\bfpage    \undefined \def\bfpage#1{#1}\fi
\ifx\blpage    \undefined \def\blpage #1{#1}\fi
\ifx\burl      \undefined \def\burl#1{\textsf{#1}}\fi
\ifx\href      \undefined \def\href#1#2{\textsf{#2}}\fi
\ifx\betal     \undefined \def\betal{\textit{et al.}}\fi
\ifx\bctitle   \undefined \def\bctitle#1{#1}\fi
\ifx\beditor   \undefined \def\beditor#1{#1}\fi
\ifx\bbtitle   \undefined \def\bbtitle#1{\textit{#1}}\fi
\ifx\bedition  \undefined \def\bedition#1{#1}\fi
\ifx\bseriesno \undefined \def\bseriesno#1{\textbf{#1}}\fi
\ifx\blocation \undefined \def\blocation#1{#1}\fi
\ifx\bsertitle \undefined \def\bsertitle#1{\textit{#1}}\fi
\ifx\bsnm      \undefined \def\bsnm#1{#1}\fi
\ifx\bsuffix   \undefined \def\bsuffix#1{#1}\fi
\ifx\bparticle \undefined \def\bparticle#1{#1}\fi
\ifx\barticle  \undefined \def\barticle#1{}\fi
\ifx\binstitute  \undefined \def\binstitute#1{#1}\fi
\ifx\bpublisher  \undefined \def\bpublisher#1{#1}\fi
\ifx\doiurl    \undefined \def\doiurl#1{\href{http://dx.doi.org/#1}{\textsf{DOI}}}\fi
\ifx\arxivurl  \undefined \def\arxivurl#1{\href{http://arxiv.org/abs/#1}{\textsf{arXiv}}}\fi
\ifx\adsurl    \undefined \def\adsurl#1{\href{http://adsabs.harvard.edu/abs/#1}{\textsf{ADS}}}\fi
\ifx\botherref \undefined \def\botherref#1{}\fi
\ifx\url       \undefined \def\url#1{\textsf{#1}}\fi
\ifx\bchapter  \undefined \def\bchapter#1{}\fi
\ifx\bbook     \undefined \def\bbook#1{}\fi
\ifx\bcomment  \undefined \def\bcomment#1{#1}\fi
\ifx\oauthor   \undefined \def\oauthor#1{#1}\fi
\ifx\citeauthoryear \undefined\def \citeauthoryear#1{#1}\fi
\ifx\endbibitem\undefined \def\endbibitem{}\fi
\ifx\bconflocation  \undefined \def\bconflocation#1{#1} \fi

\bibitem[\protect\citeauthoryear{{Abduallah} \textit{et~al.}}{2022}]{Abduallah2022ApJS}
\begin{barticle}
\bauthor{\bsnm{{Abduallah}}, \binits{Y.}},
\bauthor{\bsnm{{Jordanova}}, \binits{V.K.}},
\bauthor{\bsnm{{Liu}}, \binits{H.}},
\bauthor{\bsnm{{Li}}, \binits{Q.}},
\bauthor{\bsnm{{Wang}}, \binits{J.T.L.}},
\bauthor{\bsnm{{Wang}}, \binits{H.}}:
\byear{2022},
\batitle{Predicting solar energetic particles using {SDO/HMI} vector magnetic data products and a bidirectional {LSTM} network}.
\bjtitle{\apjs}
\bvolume{260}(\bissue{1}),
\bfpage{16}.
\doiurl{10.3847/1538-4365/ac5f56}.
\end{barticle}
\endbibitem

\bibitem[\protect\citeauthoryear{{Aschwanden}, {Xu}, and {Jing}}{2014}]{2014ApJ...797...50A}
\begin{barticle}
\bauthor{\bsnm{{Aschwanden}}, \binits{M.J.}},
\bauthor{\bsnm{{Xu}}, \binits{Y.}},
\bauthor{\bsnm{{Jing}}, \binits{J.}}:
\byear{2014},
\batitle{Global energetics of solar flares. {I. Magnetic} energies}.
\bjtitle{\apj}
\bvolume{797}(\bissue{1}),
\bfpage{50}.
\doiurl{10.1088/0004-637X/797/1/50}.
\end{barticle}
\endbibitem

\bibitem[\protect\citeauthoryear{Chen and Qi}{2018}]{8589454}
\begin{bchapter}
\bauthor{\bsnm{Chen}, \binits{C.}},
\bauthor{\bsnm{Qi}, \binits{F.}}:
\byear{2018},
\bctitle{Single image super-resolution using deep {CNN} with dense skip connections and {Inception-ResNet}}.
In: \bbtitle{2018 International Conference on Information Technology in Medicine and Education},
\bfpage{999}.
\doiurl{10.1109/ITME.2018.00222}.
\end{bchapter}
\endbibitem

\bibitem[\protect\citeauthoryear{{Deng} \textit{et~al.}}{2021}]{2021ApJ...923...76D}
\begin{barticle}
\bauthor{\bsnm{{Deng}}, \binits{J.}},
\bauthor{\bsnm{{Song}}, \binits{W.}},
\bauthor{\bsnm{{Liu}}, \binits{D.}},
\bauthor{\bsnm{{Li}}, \binits{Q.}},
\bauthor{\bsnm{{Lin}}, \binits{G.}},
\bauthor{\bsnm{{Wang}}, \binits{H.}}:
\byear{2021},
\batitle{Improving the spatial resolution of solar images using generative adversarial network and self-attention mechanism}.
\bjtitle{\apj}
\bvolume{923}(\bissue{1}),
\bfpage{76}.
\doiurl{10.3847/1538-4357/ac2aa2}.
\end{barticle}
\endbibitem

\bibitem[\protect\citeauthoryear{Deng, Hinton, and Kingsbury}{2013}]{6639344}
\begin{bchapter}
\bauthor{\bsnm{Deng}, \binits{L.}},
\bauthor{\bsnm{Hinton}, \binits{G.}},
\bauthor{\bsnm{Kingsbury}, \binits{B.}}:
\byear{2013},
\bctitle{New types of deep neural network learning for speech recognition and related applications: An overview}.
In: \bbtitle{2013 IEEE International Conference on Acoustics, Speech and Signal Processing},
\bfpage{8599}.
\doiurl{10.1109/ICASSP.2013.6639344}.
\end{bchapter}
\endbibitem

\bibitem[\protect\citeauthoryear{{D{\'\i}az Baso} and {Asensio Ramos}}{2018}]{2018A&A...614A...5D}
\begin{barticle}
\bauthor{\bsnm{{D{\'\i}az Baso}}, \binits{C.J.}},
\bauthor{\bsnm{{Asensio Ramos}}, \binits{A.}}:
\byear{2018},
\batitle{{Enhancing SDO/HMI images using deep learning}}.
\bjtitle{\aap}
\bvolume{614},
\bfpage{A5}.
\doiurl{10.1051/0004-6361/201731344}.
\end{barticle}
\endbibitem

\bibitem[\protect\citeauthoryear{{Domingo}, {Fleck}, and {Poland}}{1995}]{SOHO}
\begin{barticle}
\bauthor{\bsnm{{Domingo}}, \binits{V.}},
\bauthor{\bsnm{{Fleck}}, \binits{B.}},
\bauthor{\bsnm{{Poland}}, \binits{A.I.}}:
\byear{1995},
\batitle{The {SOHO} mission: An overview}.
\bjtitle{\solphys}
\bvolume{162}(\bissue{1-2}),
\bfpage{1}.
\doiurl{10.1007/BF00733425}.
\end{barticle}
\endbibitem

\bibitem[\protect\citeauthoryear{{Espu{\~n}a Fontcuberta} \textit{et~al.}}{2023}]{2023SoPh..298....8E}
\begin{barticle}
\bauthor{\bsnm{{Espu{\~n}a Fontcuberta}}, \binits{A.}},
\bauthor{\bsnm{{Ghosh}}, \binits{A.}},
\bauthor{\bsnm{{Chatterjee}}, \binits{S.}},
\bauthor{\bsnm{{Mitra}}, \binits{D.}},
\bauthor{\bsnm{{Nandy}}, \binits{D.}}:
\byear{2023},
\batitle{Forecasting {Solar Cycle 25} with physical model-validated recurrent neural networks}.
\bjtitle{\solphys}
\bvolume{298}(\bissue{1}),
\bfpage{8}.
\doiurl{10.1007/s11207-022-02104-3}.
\end{barticle}
\endbibitem

\bibitem[\protect\citeauthoryear{Falk \textit{et~al.}}{2019}]{Falk2019}
\begin{barticle}
\bauthor{\bsnm{Falk}, \binits{T.}},
\bauthor{\bsnm{Mai}, \binits{D.}},
\bauthor{\bsnm{Bensch}, \binits{R.}},
\bauthor{\bsnm{{\c{C}}i{\c{c}}ek}, \binits{{\"O}.}},
\bauthor{\bsnm{Abdulkadir}, \binits{A.}},
\bauthor{\bsnm{Marrakchi}, \binits{Y.}},
\bauthor{\bsnm{B{\"o}hm}, \binits{A.}},
\bauthor{\bsnm{Deubner}, \binits{J.}},
\bauthor{\bsnm{J{\"a}ckel}, \binits{Z.}},
\bauthor{\bsnm{Seiwald}, \binits{K.}},
\bauthor{\bsnm{Dovzhenko}, \binits{A.}},
\bauthor{\bsnm{Tietz}, \binits{O.}},
\bauthor{\bsnm{Dal~Bosco}, \binits{C.}},
\bauthor{\bsnm{Walsh}, \binits{S.}},
\bauthor{\bsnm{Saltukoglu}, \binits{D.}},
\bauthor{\bsnm{Tay}, \binits{T.L.}},
\bauthor{\bsnm{Prinz}, \binits{M.}},
\bauthor{\bsnm{Palme}, \binits{K.}},
\bauthor{\bsnm{Simons}, \binits{M.}},
\bauthor{\bsnm{Diester}, \binits{I.}},
\bauthor{\bsnm{Brox}, \binits{T.}},
\bauthor{\bsnm{Ronneberger}, \binits{O.}}:
\byear{2019},
\batitle{{U-Net:} deep learning for cell counting, detection, and morphometry}.
\bjtitle{Nature Methods}
\bvolume{16}(\bissue{1}),
\bfpage{67}.
\doiurl{10.1038/s41592-018-0261-2}.
\end{barticle}
\endbibitem

\bibitem[\protect\citeauthoryear{He \textit{et~al.}}{2016}]{DBLP:conf/cvpr/HeZRS16}
\begin{bchapter}
\bauthor{\bsnm{He}, \binits{K.}},
\bauthor{\bsnm{Zhang}, \binits{X.}},
\bauthor{\bsnm{Ren}, \binits{S.}},
\bauthor{\bsnm{Sun}, \binits{J.}}:
\byear{2016},
\bctitle{Deep residual learning for image recognition}.
In: \bbtitle{2016 {IEEE} Conference on Computer Vision and Pattern Recognition},
\bfpage{770}.
\doiurl{10.1109/CVPR.2016.90}.
\end{bchapter}
\endbibitem

\bibitem[\protect\citeauthoryear{Hu \textit{et~al.}}{2018}]{DBLP:journals/access/HuTPW18}
\begin{barticle}
\bauthor{\bsnm{Hu}, \binits{Z.}},
\bauthor{\bsnm{Turki}, \binits{T.}},
\bauthor{\bsnm{Phan}, \binits{N.}},
\bauthor{\bsnm{Wang}, \binits{J.T.L.}}:
\byear{2018},
\batitle{A {3D} atrous convolutional long short-term memory network for background subtraction}.
\bjtitle{{IEEE} Access}
\bvolume{6},
\bfpage{43450}.
\doiurl{10.1109/ACCESS.2018.2861223}.
\end{barticle}
\endbibitem

\bibitem[\protect\citeauthoryear{Huang and Chen}{2022}]{DBLP:journals/access/HuangC22a}
\begin{barticle}
\bauthor{\bsnm{Huang}, \binits{D.}},
\bauthor{\bsnm{Chen}, \binits{J.}}:
\byear{2022},
\batitle{{MESR:} {Multistage} enhancement network for image super-resolution}.
\bjtitle{{IEEE} Access}
\bvolume{10},
\bfpage{54599}.
\doiurl{10.1109/ACCESS.2022.3176605}.
\end{barticle}
\endbibitem

\bibitem[\protect\citeauthoryear{{Hudson}}{2011}]{2011SSRv..158....5H}
\begin{barticle}
\bauthor{\bsnm{{Hudson}}, \binits{H.S.}}:
\byear{2011},
\batitle{Global properties of solar flares}.
\bjtitle{\ssr}
\bvolume{158}(\bissue{1}),
\bfpage{5}.
\doiurl{10.1007/s11214-010-9721-4}.
\end{barticle}
\endbibitem

\bibitem[\protect\citeauthoryear{{Jiang} \textit{et~al.}}{2020}]{2020ApJS..250....5J}
\begin{barticle}
\bauthor{\bsnm{{Jiang}}, \binits{H.}},
\bauthor{\bsnm{{Wang}}, \binits{J.}},
\bauthor{\bsnm{{Liu}}, \binits{C.}},
\bauthor{\bsnm{{Jing}}, \binits{J.}},
\bauthor{\bsnm{{Liu}}, \binits{H.}},
\bauthor{\bsnm{{Wang}}, \binits{J.T.L.}},
\bauthor{\bsnm{{Wang}}, \binits{H.}}:
\byear{2020},
\batitle{Identifying and tracking solar magnetic flux elements with deep learning}.
\bjtitle{\apjs}
\bvolume{250}(\bissue{1}),
\bfpage{5}.
\doiurl{10.3847/1538-4365/aba4aa}.
\end{barticle}
\endbibitem

\bibitem[\protect\citeauthoryear{{Jiang} \textit{et~al.}}{2021}]{2021ApJS..256...20J}
\begin{barticle}
\bauthor{\bsnm{{Jiang}}, \binits{H.}},
\bauthor{\bsnm{{Jing}}, \binits{J.}},
\bauthor{\bsnm{{Wang}}, \binits{J.}},
\bauthor{\bsnm{{Liu}}, \binits{C.}},
\bauthor{\bsnm{{Li}}, \binits{Q.}},
\bauthor{\bsnm{{Xu}}, \binits{Y.}},
\bauthor{\bsnm{{Wang}}, \binits{J.T.L.}},
\bauthor{\bsnm{{Wang}}, \binits{H.}}:
\byear{2021},
\batitle{Tracing {H{\ensuremath{\alpha}}} fibrils through {Bayesian} deep learning}.
\bjtitle{\apjs}
\bvolume{256}(\bissue{1}),
\bfpage{20}.
\doiurl{10.3847/1538-4365/ac14b7}.
\end{barticle}
\endbibitem

\bibitem[\protect\citeauthoryear{{Jiang} \textit{et~al.}}{2022}]{2022ApJ...939...66J}
\begin{barticle}
\bauthor{\bsnm{{Jiang}}, \binits{H.}},
\bauthor{\bsnm{{Li}}, \binits{Q.}},
\bauthor{\bsnm{{Xu}}, \binits{Y.}},
\bauthor{\bsnm{{Hsu}}, \binits{W.}},
\bauthor{\bsnm{{Ahn}}, \binits{K.}},
\bauthor{\bsnm{{Cao}}, \binits{W.}},
\bauthor{\bsnm{{Wang}}, \binits{J.T.L.}},
\bauthor{\bsnm{{Wang}}, \binits{H.}}:
\byear{2022},
\batitle{Inferring line-of-sight velocities and doppler widths from {Stokes} profiles of {GST/NIRIS} using stacked deep neural networks}.
\bjtitle{\apj}
\bvolume{939}(\bissue{2}),
\bfpage{66}.
\doiurl{10.3847/1538-4357/ac927e}.
\end{barticle}
\endbibitem

\bibitem[\protect\citeauthoryear{Jiang \textit{et~al.}}{2023}]{JLL-2023}
\begin{barticle}
\bauthor{\bsnm{Jiang}, \binits{H.}},
\bauthor{\bsnm{Li}, \binits{Q.}},
\bauthor{\bsnm{Liu}, \binits{N.}},
\bauthor{\bsnm{Hu}, \binits{Z.}},
\bauthor{\bsnm{Abduallah}, \binits{Y.}},
\bauthor{\bsnm{Jing}, \binits{J.}},
\bauthor{\bsnm{Xu}, \binits{Y.}},
\bauthor{\bsnm{Wang}, \binits{J.T.L.}},
\bauthor{\bsnm{Wang}, \binits{H.}}:
\byear{2023},
\batitle{Generating photospheric vector magnetograms of solar active regions for {SOHO/MDI} using {SDO/HMI} and {BBSO} data with deep learning}.
\bjtitle{\solphys}
\bvolume{298},
\bfpage{87}.
\doiurl{10.1007/s11207-023-02180-z}.
\end{barticle}
\endbibitem

\bibitem[\protect\citeauthoryear{{Jonas} \textit{et~al.}}{2018}]{2018SoPh..293...48J}
\begin{barticle}
\bauthor{\bsnm{{Jonas}}, \binits{E.}},
\bauthor{\bsnm{{Bobra}}, \binits{M.}},
\bauthor{\bsnm{{Shankar}}, \binits{V.}},
\bauthor{\bsnm{{Todd Hoeksema}}, \binits{J.}},
\bauthor{\bsnm{{Recht}}, \binits{B.}}:
\byear{2018},
\batitle{Flare prediction using photospheric and coronal image data}.
\bjtitle{\solphys}
\bvolume{293}(\bissue{3}),
\bfpage{48}.
\doiurl{10.1007/s11207-018-1258-9}.
\end{barticle}
\endbibitem

\bibitem[\protect\citeauthoryear{Kastrati \textit{et~al.}}{2021}]{app11093986}
\begin{barticle}
\bauthor{\bsnm{Kastrati}, \binits{Z.}},
\bauthor{\bsnm{Dalipi}, \binits{F.}},
\bauthor{\bsnm{Imran}, \binits{A.S.}},
\bauthor{\bsnm{Pireva~Nuci}, \binits{K.}},
\bauthor{\bsnm{Wani}, \binits{M.A.}}:
\byear{2021},
\batitle{Sentiment analysis of students’ feedback with {NLP} and deep learning: A systematic mapping study}.
\bjtitle{Applied Sciences}
\bvolume{11}(\bissue{9}).
\doiurl{10.3390/app11093986}.
\end{barticle}
\endbibitem

\bibitem[\protect\citeauthoryear{Li \textit{et~al.}}{2022}]{DBLP:journals/ijon/LiYCCFXLC22}
\begin{barticle}
\bauthor{\bsnm{Li}, \binits{H.}},
\bauthor{\bsnm{Yang}, \binits{Y.}},
\bauthor{\bsnm{Chang}, \binits{M.}},
\bauthor{\bsnm{Chen}, \binits{S.}},
\bauthor{\bsnm{Feng}, \binits{H.}},
\bauthor{\bsnm{Xu}, \binits{Z.}},
\bauthor{\bsnm{Li}, \binits{Q.}},
\bauthor{\bsnm{Chen}, \binits{Y.}}:
\byear{2022},
\batitle{Srdiff: Single image super-resolution with diffusion probabilistic models}.
\bjtitle{Neurocomputing}
\bvolume{479},
\bfpage{47}.
\doiurl{10.1016/j.neucom.2022.01.029}.
\end{barticle}
\endbibitem

\bibitem[\protect\citeauthoryear{{Liu} \textit{et~al.}}{2019}]{Liu2019ApJ}
\begin{barticle}
\bauthor{\bsnm{{Liu}}, \binits{H.}},
\bauthor{\bsnm{{Liu}}, \binits{C.}},
\bauthor{\bsnm{{Wang}}, \binits{J.T.L.}},
\bauthor{\bsnm{{Wang}}, \binits{H.}}:
\byear{2019},
\batitle{Predicting solar flares using a long short-term memory network}.
\bjtitle{\apj}
\bvolume{877}(\bissue{2}),
\bfpage{121}.
\doiurl{10.3847/1538-4357/ab1b3c}.
\end{barticle}
\endbibitem

\bibitem[\protect\citeauthoryear{{Liu} \textit{et~al.}}{2020a}]{2020ApJ...894...70L}
\begin{barticle}
\bauthor{\bsnm{{Liu}}, \binits{H.}},
\bauthor{\bsnm{{Xu}}, \binits{Y.}},
\bauthor{\bsnm{{Wang}}, \binits{J.}},
\bauthor{\bsnm{{Jing}}, \binits{J.}},
\bauthor{\bsnm{{Liu}}, \binits{C.}},
\bauthor{\bsnm{{Wang}}, \binits{J.T.L.}},
\bauthor{\bsnm{{Wang}}, \binits{H.}}:
\byear{2020}a,
\batitle{Inferring vector magnetic fields from {Stokes} profiles of {GST/NIRIS} using a convolutional neural network}.
\bjtitle{\apj}
\bvolume{894}(\bissue{1}),
\bfpage{70}.
\doiurl{10.3847/1538-4357/ab8818}.
\end{barticle}
\endbibitem

\bibitem[\protect\citeauthoryear{{Liu} \textit{et~al.}}{2020b}]{Liu2020ApJ}
\begin{barticle}
\bauthor{\bsnm{{Liu}}, \binits{H.}},
\bauthor{\bsnm{{Liu}}, \binits{C.}},
\bauthor{\bsnm{{Wang}}, \binits{J.T.L.}},
\bauthor{\bsnm{{Wang}}, \binits{H.}}:
\byear{2020}b,
\batitle{Predicting coronal mass ejections using {SDO/HMI} vector magnetic data products and recurrent neural networks}.
\bjtitle{\apj}
\bvolume{890}(\bissue{1}),
\bfpage{12}.
\doiurl{10.3847/1538-4357/ab6850}.
\end{barticle}
\endbibitem

\bibitem[\protect\citeauthoryear{{Liu} \textit{et~al.}}{2022}]{2022ApJ...941...20L}
\begin{barticle}
\bauthor{\bsnm{{Liu}}, \binits{S.}},
\bauthor{\bsnm{{Xu}}, \binits{L.}},
\bauthor{\bsnm{{Zhao}}, \binits{Z.}},
\bauthor{\bsnm{{Erd{\'e}lyi}}, \binits{R.}},
\bauthor{\bsnm{{Kors{\'o}s}}, \binits{M.B.}},
\bauthor{\bsnm{{Huang}}, \binits{X.}}:
\byear{2022},
\batitle{Deep learning based solar flare forecasting model. {II.} {Influence} of image resolution}.
\bjtitle{\apj}
\bvolume{941}(\bissue{1}),
\bfpage{20}.
\doiurl{10.3847/1538-4357/ac99dc}.
\end{barticle}
\endbibitem

\bibitem[\protect\citeauthoryear{{Liu} \textit{et~al.}}{2012}]{2012SoPh..279..295L}
\begin{barticle}
\bauthor{\bsnm{{Liu}}, \binits{Y.}},
\bauthor{\bsnm{{Hoeksema}}, \binits{J.T.}},
\bauthor{\bsnm{{Scherrer}}, \binits{P.H.}},
\bauthor{\bsnm{{Schou}}, \binits{J.}},
\bauthor{\bsnm{{Couvidat}}, \binits{S.}},
\bauthor{\bsnm{{Bush}}, \binits{R.I.}},
\bauthor{\bsnm{{Duvall}}, \binits{T.L.}},
\bauthor{\bsnm{{Hayashi}}, \binits{K.}},
\bauthor{\bsnm{{Sun}}, \binits{X.}},
\bauthor{\bsnm{{Zhao}}, \binits{X.}}:
\byear{2012},
\batitle{Comparison of line-of-sight magnetograms taken by the {Solar Dynamics Observatory/Helioseismic and Magnetic Imager} and {Solar and Heliospheric Observatory/Michelson Doppler Imager}}.
\bjtitle{\solphys}
\bvolume{279}(\bissue{1}),
\bfpage{295}.
\doiurl{10.1007/s11207-012-9976-x}.
\end{barticle}
\endbibitem

\bibitem[\protect\citeauthoryear{{Mayfield} and {Lawrence}}{1985}]{1985SoPh...96..293M}
\begin{barticle}
\bauthor{\bsnm{{Mayfield}}, \binits{E.B.}},
\bauthor{\bsnm{{Lawrence}}, \binits{J.K.}}:
\byear{1985},
\batitle{The correlation of solar flare production with magnetic energy in active regions}.
\bjtitle{\solphys}
\bvolume{96}(\bissue{2}),
\bfpage{293}.
\doiurl{10.1007/BF00149685}.
\end{barticle}
\endbibitem

\bibitem[\protect\citeauthoryear{{Mercea} \textit{et~al.}}{2023}]{2023SoPh..298....4M}
\begin{barticle}
\bauthor{\bsnm{{Mercea}}, \binits{V.}},
\bauthor{\bsnm{{Paraschiv}}, \binits{A.R.}},
\bauthor{\bsnm{{Lacatus}}, \binits{D.A.}},
\bauthor{\bsnm{{Marginean}}, \binits{A.}},
\bauthor{\bsnm{{Besliu-Ionescu}}, \binits{D.}}:
\byear{2023},
\batitle{A machine learning enhanced approach for automated sunquake detection in acoustic emission maps}.
\bjtitle{\solphys}
\bvolume{298}(\bissue{1}),
\bfpage{4}.
\doiurl{10.1007/s11207-022-02081-7}.
\end{barticle}
\endbibitem

\bibitem[\protect\citeauthoryear{Misra}{2020}]{DBLP:conf/bmvc/Misra20}
\begin{bchapter}
\bauthor{\bsnm{Misra}, \binits{D.}}:
\byear{2020},
\bctitle{Mish: {A} self regularized non-monotonic activation function}.
In: \bbtitle{31st British Machine Vision Conference}.
\burl{https://www.bmvc2020-conference.com/assets/papers/0928.pdf}.
\end{bchapter}
\endbibitem

\bibitem[\protect\citeauthoryear{{Pesnell}, {Thompson}, and {Chamberlin}}{2012}]{SDO}
\begin{barticle}
\bauthor{\bsnm{{Pesnell}}, \binits{W.D.}},
\bauthor{\bsnm{{Thompson}}, \binits{B.J.}},
\bauthor{\bsnm{{Chamberlin}}, \binits{P.C.}}:
\byear{2012},
\batitle{{The Solar Dynamics Observatory (SDO)}}.
\bjtitle{\solphys}
\bvolume{275},
\bfpage{3}.
\doiurl{10.1007/s11207-011-9841-3}.
\end{barticle}
\endbibitem

\bibitem[\protect\citeauthoryear{{Priest}, {Longcope}, and {Janvier}}{2016}]{2016SoPh..291.2017P}
\begin{barticle}
\bauthor{\bsnm{{Priest}}, \binits{E.R.}},
\bauthor{\bsnm{{Longcope}}, \binits{D.W.}},
\bauthor{\bsnm{{Janvier}}, \binits{M.}}:
\byear{2016},
\batitle{Evolution of magnetic helicity during eruptive flares and coronal mass ejections}.
\bjtitle{\solphys}
\bvolume{291}(\bissue{7}),
\bfpage{2017}.
\doiurl{10.1007/s11207-016-0962-6}.
\end{barticle}
\endbibitem

\bibitem[\protect\citeauthoryear{Qin \textit{et~al.}}{2021}]{DBLP:conf/iccv/QinZW021}
\begin{bchapter}
\bauthor{\bsnm{Qin}, \binits{Z.}},
\bauthor{\bsnm{Zhang}, \binits{P.}},
\bauthor{\bsnm{Wu}, \binits{F.}},
\bauthor{\bsnm{Li}, \binits{X.}}:
\byear{2021},
\bctitle{{FcaNet}: Frequency channel attention networks}.
In: \bbtitle{2021 {IEEE/CVF} International Conference on Computer Vision},
\bfpage{763}.
\doiurl{10.1109/ICCV48922.2021.00082}.
\end{bchapter}
\endbibitem

\bibitem[\protect\citeauthoryear{Rahim, Hassan, and Shin}{2021}]{DBLP:journals/bspc/RahimHS21}
\begin{barticle}
\bauthor{\bsnm{Rahim}, \binits{T.}},
\bauthor{\bsnm{Hassan}, \binits{S.A.}},
\bauthor{\bsnm{Shin}, \binits{S.Y.}}:
\byear{2021},
\batitle{A deep convolutional neural network for the detection of polyps in colonoscopy images}.
\bjtitle{Biomed. Signal Process. Control.}
\bvolume{68},
\bfpage{102654}.
\doiurl{10.1016/j.bspc.2021.102654}.
\end{barticle}
\endbibitem

\bibitem[\protect\citeauthoryear{{Rahman} \textit{et~al.}}{2020}]{2020ApJ...897L..32R}
\begin{barticle}
\bauthor{\bsnm{{Rahman}}, \binits{S.}},
\bauthor{\bsnm{{Moon}}, \binits{Y.-J.}},
\bauthor{\bsnm{{Park}}, \binits{E.}},
\bauthor{\bsnm{{Siddique}}, \binits{A.}},
\bauthor{\bsnm{{Cho}}, \binits{I.-H.}},
\bauthor{\bsnm{{Lim}}, \binits{D.}}:
\byear{2020},
\batitle{Super-resolution of {SDO/HMI} magnetograms using novel deep learning methods}.
\bjtitle{\apjl}
\bvolume{897}(\bissue{2}),
\bfpage{L32}.
\doiurl{10.3847/2041-8213/ab9d79}.
\end{barticle}
\endbibitem

\bibitem[\protect\citeauthoryear{{Reames}}{2022}]{2022SoPh..297...32R}
\begin{barticle}
\bauthor{\bsnm{{Reames}}, \binits{D.V.}}:
\byear{2022},
\batitle{Energy spectra vs. element abundances in solar energetic particles and the roles of magnetic reconnection and shock acceleration}.
\bjtitle{\solphys}
\bvolume{297}(\bissue{3}),
\bfpage{32}.
\doiurl{10.1007/s11207-022-01961-2}.
\end{barticle}
\endbibitem

\bibitem[\protect\citeauthoryear{Sara, Akter, and Uddin}{2019}]{sara2019image}
\begin{barticle}
\bauthor{\bsnm{Sara}, \binits{U.}},
\bauthor{\bsnm{Akter}, \binits{M.}},
\bauthor{\bsnm{Uddin}, \binits{M.S.}}:
\byear{2019},
\batitle{Image quality assessment through {FSIM}, {SSIM}, {MSE} and {PSNR}—{A} comparative study}.
\bjtitle{Journal of Computer and Communications}
\bvolume{7}(\bissue{3}),
\bfpage{8}.
\doiurl{10.4236/jcc.2019.73002}.
\end{barticle}
\endbibitem

\bibitem[\protect\citeauthoryear{{Scherrer} \textit{et~al.}}{1995}]{MDI}
\begin{barticle}
\bauthor{\bsnm{{Scherrer}}, \binits{P.H.}},
\bauthor{\bsnm{{Bogart}}, \binits{R.S.}},
\bauthor{\bsnm{{Bush}}, \binits{R.I.}},
\bauthor{\bsnm{{Hoeksema}}, \binits{J.T.}},
\bauthor{\bsnm{{Kosovichev}}, \binits{A.G.}},
\bauthor{\bsnm{{Schou}}, \binits{J.}},
\bauthor{\bsnm{{Rosenberg}}, \binits{W.}},
\bauthor{\bsnm{{Springer}}, \binits{L.}},
\bauthor{\bsnm{{Tarbell}}, \binits{T.D.}},
\bauthor{\bsnm{{Title}}, \binits{A.}},
\bauthor{\bsnm{{Wolfson}}, \binits{C.J.}},
\bauthor{\bsnm{{Zayer}}, \binits{I.}},
\bauthor{\bsnm{{The MDI Engineering Team}}}:
\byear{1995},
\batitle{{The Solar Oscillations Investigation - Michelson Doppler Imager}}.
\bjtitle{\solphys}
\bvolume{162},
\bfpage{129}.
\doiurl{10.1007/BF00733429}.
\end{barticle}
\endbibitem

\bibitem[\protect\citeauthoryear{{Schou} \textit{et~al.}}{2012}]{HMI}
\begin{barticle}
\bauthor{\bsnm{{Schou}}, \binits{J.}},
\bauthor{\bsnm{{Scherrer}}, \binits{P.H.}},
\bauthor{\bsnm{{Bush}}, \binits{R.I.}},
\bauthor{\bsnm{{Wachter}}, \binits{R.}},
\bauthor{\bsnm{{Couvidat}}, \binits{S.}},
\bauthor{\bsnm{{Rabello-Soares}}, \binits{M.C.}},
\bauthor{\bsnm{{Bogart}}, \binits{R.S.}},
\bauthor{\bsnm{{Hoeksema}}, \binits{J.T.}},
\bauthor{\bsnm{{Liu}}, \binits{Y.}},
\bauthor{\bsnm{{Duvall}}, \binits{T.L.}},
\bauthor{\bsnm{{Akin}}, \binits{D.J.}},
\bauthor{\bsnm{{Allard}}, \binits{B.A.}},
\bauthor{\bsnm{{Miles}}, \binits{J.W.}},
\bauthor{\bsnm{{Rairden}}, \binits{R.}},
\bauthor{\bsnm{{Shine}}, \binits{R.A.}},
\bauthor{\bsnm{{Tarbell}}, \binits{T.D.}},
\bauthor{\bsnm{{Title}}, \binits{A.M.}},
\bauthor{\bsnm{{Wolfson}}, \binits{C.J.}},
\bauthor{\bsnm{{Elmore}}, \binits{D.F.}},
\bauthor{\bsnm{{Norton}}, \binits{A.A.}},
\bauthor{\bsnm{{Tomczyk}}, \binits{S.}}:
\byear{2012},
\batitle{Design and ground calibration of the {Helioseismic and Magnetic Imager} {(HMI)} instrument on the {Solar Dynamics Observatory} {(SDO)}}.
\bjtitle{\solphys}
\bvolume{275},
\bfpage{229}.
\doiurl{10.1007/s11207-011-9842-2}.
\end{barticle}
\endbibitem

\bibitem[\protect\citeauthoryear{{Scully} \textit{et~al.}}{2023}]{2023SoPh..298....6S}
\begin{barticle}
\bauthor{\bsnm{{Scully}}, \binits{J.}},
\bauthor{\bsnm{{Flynn}}, \binits{R.}},
\bauthor{\bsnm{{Carley}}, \binits{E.}},
\bauthor{\bsnm{{Gallagher}}, \binits{P.}},
\bauthor{\bsnm{{Daly}}, \binits{M.}}:
\byear{2023},
\batitle{Simulating solar radio bursts using generative adversarial networks}.
\bjtitle{\solphys}
\bvolume{298}(\bissue{1}),
\bfpage{6}.
\doiurl{10.1007/s11207-022-02099-x}.
\end{barticle}
\endbibitem

\bibitem[\protect\citeauthoryear{{Song} \textit{et~al.}}{2022}]{2022ApJS..263...25S}
\begin{barticle}
\bauthor{\bsnm{{Song}}, \binits{W.}},
\bauthor{\bsnm{{Ma}}, \binits{W.}},
\bauthor{\bsnm{{Ma}}, \binits{Y.}},
\bauthor{\bsnm{{Zhao}}, \binits{X.}},
\bauthor{\bsnm{{Lin}}, \binits{G.}}:
\byear{2022},
\batitle{Improving the spatial resolution of solar images based on an improved conditional denoising diffusion probability model}.
\bjtitle{\apjs}
\bvolume{263}(\bissue{2}),
\bfpage{25}.
\doiurl{10.3847/1538-4365/ac9a4d}.
\end{barticle}
\endbibitem

\bibitem[\protect\citeauthoryear{{Tsuneta} \textit{et~al.}}{2008}]{2008SoPh..249..167T}
\begin{barticle}
\bauthor{\bsnm{{Tsuneta}}, \binits{S.}},
\bauthor{\bsnm{{Ichimoto}}, \binits{K.}},
\bauthor{\bsnm{{Katsukawa}}, \binits{Y.}},
\bauthor{\bsnm{{Nagata}}, \binits{S.}},
\bauthor{\bsnm{{Otsubo}}, \binits{M.}},
\bauthor{\bsnm{{Shimizu}}, \binits{T.}},
\bauthor{\bsnm{{Suematsu}}, \binits{Y.}},
\bauthor{\bsnm{{Nakagiri}}, \binits{M.}},
\bauthor{\bsnm{{Noguchi}}, \binits{M.}},
\bauthor{\bsnm{{Tarbell}}, \binits{T.}},
\bauthor{\bsnm{{Title}}, \binits{A.}},
\bauthor{\bsnm{{Shine}}, \binits{R.}},
\bauthor{\bsnm{{Rosenberg}}, \binits{W.}},
\bauthor{\bsnm{{Hoffmann}}, \binits{C.}},
\bauthor{\bsnm{{Jurcevich}}, \binits{B.}},
\bauthor{\bsnm{{Kushner}}, \binits{G.}},
\bauthor{\bsnm{{Levay}}, \binits{M.}},
\bauthor{\bsnm{{Lites}}, \binits{B.}},
\bauthor{\bsnm{{Elmore}}, \binits{D.}},
\bauthor{\bsnm{{Matsushita}}, \binits{T.}},
\bauthor{\bsnm{{Kawaguchi}}, \binits{N.}},
\bauthor{\bsnm{{Saito}}, \binits{H.}},
\bauthor{\bsnm{{Mikami}}, \binits{I.}},
\bauthor{\bsnm{{Hill}}, \binits{L.D.}},
\bauthor{\bsnm{{Owens}}, \binits{J.K.}}:
\byear{2008},
\batitle{The {Solar Optical Telescope} for the {Hinode} mission: An overview}.
\bjtitle{\solphys}
\bvolume{249}(\bissue{2}),
\bfpage{167}.
\doiurl{10.1007/s11207-008-9174-z}.
\end{barticle}
\endbibitem

\bibitem[\protect\citeauthoryear{Webb and Howard}{2012}]{webb2012coronal}
\begin{barticle}
\bauthor{\bsnm{Webb}, \binits{D.F.}},
\bauthor{\bsnm{Howard}, \binits{T.A.}}:
\byear{2012},
\batitle{Coronal mass ejections: Observations}.
\bjtitle{Living Rev. Solar Phys.}
\bvolume{9}(\bissue{1}),
\bfpage{1}.
\doiurl{10.12942/lrsp-2012-3}.
\end{barticle}
\endbibitem

\bibitem[\protect\citeauthoryear{{Wedemeyer-B{\"o}hm} and {Rouppe van der Voort}}{2009}]{2009A&A...503..225W}
\begin{barticle}
\bauthor{\bsnm{{Wedemeyer-B{\"o}hm}}, \binits{S.}},
\bauthor{\bsnm{{Rouppe van der Voort}}, \binits{L.}}:
\byear{2009},
\batitle{{On the continuum intensity distribution of the solar photosphere}}.
\bjtitle{\aap}
\bvolume{503}(\bissue{1}),
\bfpage{225}.
\doiurl{10.1051/0004-6361/200911983}.
\end{barticle}
\endbibitem

\bibitem[\protect\citeauthoryear{Yang \textit{et~al.}}{2019}]{DBLP:journals/tmm/YangZTWXL19}
\begin{barticle}
\bauthor{\bsnm{Yang}, \binits{W.}},
\bauthor{\bsnm{Zhang}, \binits{X.}},
\bauthor{\bsnm{Tian}, \binits{Y.}},
\bauthor{\bsnm{Wang}, \binits{W.}},
\bauthor{\bsnm{Xue}, \binits{J.}},
\bauthor{\bsnm{Liao}, \binits{Q.}}:
\byear{2019},
\batitle{Deep learning for single image super-resolution: {A} brief review}.
\bjtitle{{IEEE} Trans. Multim.}
\bvolume{21}(\bissue{12}),
\bfpage{3106}.
\doiurl{10.1109/TMM.2019.2919431}.
\end{barticle}
\endbibitem

\bibitem[\protect\citeauthoryear{{Zhu} \textit{et~al.}}{2019}]{2019SoPh..294..117Z}
\begin{barticle}
\bauthor{\bsnm{{Zhu}}, \binits{G.}},
\bauthor{\bsnm{{Lin}}, \binits{G.}},
\bauthor{\bsnm{{Wang}}, \binits{D.}},
\bauthor{\bsnm{{Liu}}, \binits{S.}},
\bauthor{\bsnm{{Yang}}, \binits{X.}}:
\byear{2019},
\batitle{Solar filament recognition based on deep learning}.
\bjtitle{\solphys}
\bvolume{294}(\bissue{9}),
\bfpage{117}.
\doiurl{10.1007/s11207-019-1517-4}.
\end{barticle}
\endbibitem

\end{thebibliography}

\end{article} 

\end{document}